\documentclass{ws-ijgmmp}
\usepackage{color}

\begin{document}

\markboth{Kyosuke TOMONARI}
{A unified-description of curvature, torsion, and non-metricity}

%
\catchline{}{}{}{}{}
%

\title{A unified-description of curvature, torsion, and non-metricity of the metric-affine geometry with the M\"{o}bius representation}

\author{Kyosuke TOMONARI}
\address{Department of Physics, Tokyo Institute of Technology,\\
2-12-1 Ookayama, Meguro-ku, Tokyo 152-8551, Japan.
\email{ktomonari.phys@gmail.com}}

\maketitle

\begin{history}
\received{(Day Month Year)}
\revised{(Day Month Year)}
\end{history}

\begin{abstract}
We establish the mathematical fundamentals for a unified description of curvature, torsion, and non-metricity 2-forms in the way extending the so-called M\"{o}bius representation of the affine group, which is the method to convert the semi-direct product into the ordinary matrix product, to revive the fertility of gauge theories of gravity. First of all, we illustrate the basic concepts for constructing the metric-affine geometry. Then the curvature and torsion 2-forms are described in a unified manner by using the Cartan connection of the M\"{o}bius representation of the affine group. In this unified-description, the curvature and torsion are derived by Cartan's structure equation with respect to a common connection 1-form. After that, extending the M\"{o}bius representation, the dilation and shear 2-forms, or equivalently, the non-metricity 2-form, are introduced in the same unified manner. Based on the unified-description established in this paper, introducing a new group parametrization and applying the In\"{o}n\"{u}-Wigner group contraction to the full theory, the relationships among symmetries, geometric quantities, and geometries are investigated with respect to the three gauge groups: the metric-affine group and its extension, and an extension of the (anti)-de Sitter group in which the non-metricity exists. Finally, possible applications to theories of gravity are briefly discussed. 
\end{abstract}


\section{\label{01}Introduction}
General Relativity (GR), which is, on one hand, the most successful theory of gravity, employs the (pseudo-)Riemannian geometry to describe gravitational phenomena based on Einstein's equivalence principle and the general covariance~\cite{Einstein:1916vd,Wald:1984rg}. On the other hand, Yang-Mills gauge field theory, which describes fundamental interactions in the elementary particle physics such as electromagnetic, weak, and strong interactions in a unified manner based on the gauge principle, makes use of a principal bundle with a specific gauge group for each theory~\cite{Weyl:1929fm,Yang:1954,Weinberg:1967,Salam:1968rm,Greenberg:1964pe,Han:1965pf}. Inspired by these studies, R. Utiyama tried to establish a unified-description of all the above phenomena that is disciplined by the gauge invariant characteristics~\cite{Utiyama:1956sy}. This is the dawn of the gauge theories of gravity. A detailed and historical development of this approach is illustrated in Ref.~\cite{Blagojevic:2013xpa}. 

Metric-Affine gauge theories of Gravity (MAG) are one of such approach to gauge theories of gravity~\cite{Hehl:1986ei,Hehl:1989ij,Hehl:1994ue}. This theory is constructed upon the metric-affine geometry of which geometric quantities are not only curvature but also torsion and non-metricity (the latter one is further decomposed into dilation and shear). Therefore, MAG is an extension/modification of GR in terms of the gauge theories. In particular, as a modification of GR, a set of subclasses of MAG labeled as the Geometrical Trinity of Gravity (GTG) has been vigorously investigated by virtue of its equivalence in the equations of motion to GR up to boundary terms in the meaning of the action functional~\cite{BeltranJimenez:2019esp}. Under the imposition of the so-called Weitzenb\"{o}ch gauge condition~\cite{Weitzenboch:1923,Hayashi:1979,Hayashi:1981,Blagojevic:2020dyq,BeltranJimenez:2022azb}, which implies that the curvature vanishes and then the geometry is flat, GTG has two subclasses: Symmetric Teleparallel Equivalent to GR (STEGR)~\cite{Nester:1998mp} and Teleparallel Equivalent to GR (TEGR)~\cite{Einstein:1928}. 

STEGR is, on one hand, obtained by the imposition of vanishing torsion that is realized in the use of the so-called St\"{u}kelberg fields~\cite{BeltranJimenez:2019esp}. This theory describes gravity in terms only of the non-metricity. Furthermore, the imposition of the so-called coincident gauge leads to a specific subclass that is equivalent to GR with excepting the second-order derivative terms of its action functional~\cite{BeltranJimenez:2017tkd,BeltranJimenez:2022azb}. It is nothing but what Einstein first treated for deriving his field equations when applying the variational principle~\cite{Einstein:1916cd}, although it implies also that this subclass of GTG is deprived of the property of the covariance~\cite{Mukhopadhyay:2006vu}. On the other hand, TEGR is obtained by the imposition of vanishing non-metricity and describes gravity in terms only of the torsion. However, differing from STEGR, unveiling the true identity of vanishing non-metricity from the viewpoint of its gauge structure is still under development. One of the reasons for this situation would be the lack of a unified-description of non-metricity with curvature and torsion. Here, the jargon the ``{\it unified-description}'', this is nothing but the main subject of the current paper, means that both curvature and non-metricity can be derived from Cartan's structure equation by using a common connection 1-form. In fact, geometries that are characterized by both curvature and torsion have the unified-description labeled as Riemann-Cartan geometry~\cite{Cartan:1951,Kobayashi:1989,Kobayashi:1996}. In this geometry, these geometric quantities are constructed by the so-called Cartan connection, which is the extension of the Ehresmann connection by using the M\"{o}bius representation of a principal bundle with the affine gauge group~\cite{Kobayashi:1972,Mielke:1987,Hehl:1994ue}. This approach to constructing the geometries makes it possible to unveil its gauge structure and the true identity of vanishing torsion (and also vanishing curvature)~\cite{BeltranJimenez:2022azb,Gomes:2023hyk,Bahamonde:2023piz,Bahamonde:2024sqo}. Therefore, a unified-description in the same manner that takes also non-metricity into account would open the way for unveiling the true identity of vanishing non-metricity. However, to author's knowledge, there is no previous work for the unified-description on curvature and non-metricity. In addition to this point, once this approach is realized, the unified-description would make it possible to scrutinize the gauge structure not only of GTG but also of MAG. Furthermore, just alternating the gauge group, the unified-description allows us to easily switch the metric-affine geometry to another geometry. Combining the so-called In\"{o}n\"{u}-Wigner group contraction~\cite{Segal:1951,Wigner:1953}, the relationships among each geometry could be unveiled, and a new geometry, or equivalently, a new geometric quantity, would also be discovered. The purpose of the current paper is to provide the mathematical fundamentals of a unified-description of curvature, torsion, and non-metricity for the sake of the above investigations.

The paper is organized as follows: In Sec.~\ref{03}, the concept of internal space bundle is introduced, in which gauge structures for gravity will be engraved. Frame fields/vielbein are also introduced as a bundle morphism between the internal space bundle and the tangent bundle of a differential manifold. This bundle morphism will attribute the gauge structures to gravity. In Sec.~\ref{04}, based on the so-called M\"{o}bius representation, which is a method to compute the semi-direct product using the ordinary matrix product, the curvature and torsion 2-forms are reformulated by making use of the so-called Cartan connection. Then possible geometries are classified. In Sec.~\ref{05} and Sec.~\ref{06}, extending the M\"{o}bius representation of the affine group and the Cartan connection, not only curvature and torsion 2-forms but also dilation and shear 2-forms, or equivalently, non-metricity 2-form, are derived from Cartan's structure equation based on a common connection 1-form. This is nothing but the unified-description proposed in the current paper. Then possible geometries are classified. In Sec.~\ref{07}, introducing a new parametrization of the metric-affine group and using the In\"{o}n\"{u}-Wigner group contraction, the correspondences between each geometry and algebra are clarified. It will be unveiled that dilation and shear, or equivalently, non-metricity, drop out by performing the contraction only under the imposition of appropriate gauge conditions. Curvature and torsion demand also appropriate gauge conditions for vanishing in the contractions. In addition, the cases of the other two gauge groups, the extensions of the metric-affine group and the de Sitter/anti-de Sitter group, are investigated. It will be shown that new geometric quantities arise. Finally, in Sec.~\ref{08}, possible applications to theories of gravity are briefly mentioned. 

\section{\label{03}Basic ingredients in gauge theories of gravity}
In this section, a set of fundamental concepts for constructing the gauge theories of gravity is introduced. Gauge structures are engraved into an internal space (bundle), and frame fields/vielbein connect the gauge structures in the internal space to the geometric quantities in the tangent bundle of a differential manifold. In particular, in physics, it should be a spacetime manifold. An introduction to the metric-affine geometry based on bundle theory is reviewed in~\ref{02}.

\subsection{\label{03:01}Frame fields, internal space bundle, and Weitzenb\"{o}ch connection}
Frame fields/vielbein are the components of a bundle morphism~\cite{Nakahara:2003nw,Baez:1995sj}, denote $\mathbf{e}$, from a vector bundle $\mathfrak{V}=(\mathcal{V}\,,\mathcal{M}\,,\pi)$, where $\mathcal{V}$ is a total space, $\mathcal{M}$ is a base manifold, and $\pi$ is the projection map, to the tangent bundle of the manifold $\mathfrak{T}=(T\mathcal{M}\,,\mathcal{M}\,,\iota)$. That is, for any open set $U$ of an open covering of $\mathcal{M}$ and a smooth function $f\,:\,M\rightarrow M$, the bundle map $\mathbf{e}=\pi^{-1}\circ f\circ\iota\,:\,\mathcal{V}\rightarrow T\mathcal{M}\,;\,\mathcal{V}|_{U}\mapsto \mathbf{e}(\mathcal{V}|_{U})=T\mathcal{M}|_{U}$ defines a set of sections as follows:
\begin{equation}
    e_{I}=\mathbf{e}(\xi_{I})=e_{I}{}^{\mu}\partial_{\mu}\,,
\label{}
\end{equation}
where $\xi_{I}$ and $\partial_{\mu}$ are a basis of sections of $\mathcal{V}|_{U}$ and the local coordinate basis of $T\mathcal{M}|_{U}$, respectively. Then the set of components $e_{I}{}^{\mu}$, or equivalently, $e_{I}$, and the vector bundle $\mathfrak{V}$ are called frame fields/vielbein with respect to the vector bundle $\mathfrak{V}$ and the internal space bundle, respectively. Remark that both $e_{I}$ and $e_{I}{}^{\mu}$ are sections of the tangent bundle $T\mathcal{M}$, although the indices represent that of the internal space bundle $\mathcal{V}$. Therefore, the bundle morphism can also be written as follows~\cite{Carroll:2004st}:
\begin{equation}
    \mathbf{e}=e_{I}{}^{\mu}\partial_{\mu}\otimes\xi^{I}\,,
\label{Frame field as a section of TM Otimes dualV}
\end{equation}
where $\xi^{I}$ are the dual basis of $\xi_{I}$. This is a section of $T\mathcal{M}\otimes \mathcal{V}^{*}$. The projection map of the constructed bundle is obtained just by taking the tensor product of each projection map in its order. So are the local trivializations.

The inverse map of $\mathbf{e}$ can be introduced in a well-defined manner by virtue of its definition $\mathbf{e}=\pi^{-1}\circ f\circ\iota$. That is, $\mathbf{e}^{-1}=\iota^{-1}\circ f^{-1}\circ\pi\,:\,T\mathcal{M}\rightarrow\mathcal{V}\,;\,T\mathcal{M}|_{U}\mapsto \mathbf{e}^{-1}(T\mathcal{M}|_{U})=\mathcal{V}|_{U}$. Using this map, the dual basis of $\xi_{I}$, {\it i\,.e\,.,} $\xi^{I}$, are pulled back to $T^{*}\mathcal{M}$ as follows:
\begin{equation}
    e^{I}=(\mathbf{e}^{-1})^{*}(\xi^{I})=e^{I}{}_{\mu}dx^{\mu}\,,
\label{}
\end{equation}
where $dx^{\mu}$ are the dual basis of $\partial_{\mu}$. Then the components $e^{I}{}_{\mu}$, or equivalently, $e^{I}$, are called co-frame fields/co-vielbein with respect to the internal space bundle $\mathfrak{V}$. Remark that both $e^{I}$ and $e^{I}{}_{\mu}$ are sections of the co-tangent vector bundle $T^{*}\mathcal{M}$. Therefore, the inverse map can be expressed as follows~\cite{Carroll:2004st}:
\begin{equation}
    (\mathbf{e}^{-1})^{*}=e^{I}{}_{\mu}dx^{\mu}\otimes\xi_{I}\,.
\label{Frame field as a section of dualTM Otimes V}
\end{equation}
This is a section of the dual vector bundle $T^{*}\mathcal{M}\otimes\mathcal{V}$ of $T\mathcal{M}\otimes\mathcal{V}^{*}$. The projection map and local trivializations are also introduced in the dual manner.  

The frame fields/co-frame fields satisfy the following relations~\cite{Wald:1984rg,Carroll:2004st}:
\begin{equation}
    e_{I}{}^{\mu}e^{J}{}_{\mu}=\delta_{I}^{J}\,,\quad e_{I}{}^{\mu}e^{I}{}_{\nu}=\delta_{\nu}^{\mu}\,.
\label{Veilbein contranction}
\end{equation}
For a metric tensor of the tangent space $g=g_{\mu\nu}dx^{\mu}\otimes dx^{\nu}$ and a metric of the internal space bundle $g=g_{IJ}\xi^{I}\otimes \xi^{J}$, pulling back it by $\mathbf{e}$ and $(\mathbf{e}^{-1})^{*}$, the following relation is obtained~\cite{Wald:1984rg,Carroll:2004st}:
\begin{equation}
    g_{IJ}=e_{I}{}^{\mu}e_{J}{}^{\nu}g_{\mu\nu}\,,\quad g_{\mu\nu}=e^{I}{}_{\mu}e^{J}{}_{\nu}g_{IJ}\,,
\label{Metric-Vielbein relation}
\end{equation}
respectively. This relation connects the metric tensor on the internal space to that on the manifold, and vice versa. 

Since the frame field $\mathbf{e}$ can be expressed as Eq.~(\ref{Frame field as a section of TM Otimes dualV}), {\it i\,.e\,.,} a section of $T^{*}\mathcal{M}\otimes\mathcal{V}$, the covariant derivative of $\mathbf{e}$, denote $\mathcal{D}\mathbf{e}$, becomes as follows:
\begin{equation}
    \mathcal{D}\mathbf{e}=(de^{I}{}_{\mu}-\tilde{\Gamma}^{\rho}_{\nu\mu}e^{I}{}_{\rho}dx^{\nu}+\omega^{I}{}_{J\mu}e^{J}{}_{\nu}dx^{\nu})dx^{\mu}\otimes\xi_{I}\,,
\label{}
\end{equation}
where $\tilde{\Gamma}^{\rho}_{\mu\nu}$ and $\omega^{I}{}_{J\mu}$ are the affine connection components of the tangent bundle and the connection 1-form components of the internal space bundle, respectively. In component form, the above formula is expressed as follows:
\begin{equation}
    \mathcal{D}_{\mu}e^{I}{}_{\nu}=\partial_{\mu}e^{I}{}_{\nu}-\tilde{\Gamma}^{\rho}_{\mu\nu}e^{I}{}_{\rho}+\omega^{I}{}_{J\mu}e^{J}{}_{\nu}\,.
\label{}
\end{equation}
In the same way, $\mathcal{D}_{\mu}e_{I}{}^{\nu}$ is calculated as follows:
\begin{equation}
    \mathcal{D}_{\mu}e_{I}{}^{\nu}=\partial_{\mu}e_{I}{}^{\nu}+\tilde{\Gamma}^{\nu}_{\mu\rho}e_{I}{}^{\rho}-\omega^{J}{}_{I\mu}e_{J}{}^{\nu}\,.
\label{}
\end{equation}
These formulae are result from $\mathcal{V}|_{U}\simeq T\mathcal{M}|_{U}$ on a local region $U$ in $\mathcal{M}$; this local relation implies that the affine connection and the connection 1-form can be identified in the local region $U$. The transition functions $\tau_{ij}=\varphi^{-1}_{i}\circ\varphi_{j}$ of the internal space bundle $\mathcal{V}$ form its structure group $G$.~\footnote{
In particular, if $G$ is the Poincare group the connection 1-form results in the so-called spin connection.
} Therefore, for $\Lambda\in G$, the gauge transformation law of Ehressmann connection (See Eq.~(\ref{Algebraic Ehresmann connection}) in detail) leads to the transformation of the connection 1-form as follows:
\begin{equation}
    \omega^{I}{}_{J\mu}\rightarrow\omega'^{I}{}_{J\mu}=(\Lambda^{-1})^{I}{}_{K}\partial_{\mu}\Lambda^{K}{}_{J}+(\Lambda^{-1})^{I}{}_{K}\omega^{K}{}_{L\mu}\Lambda^{L}{}_{J}\,.
\label{Gauge transformation of spin connection}
\end{equation}
Then, transforming the co-frame fields $e^{I}{}_{\mu}$ as $e^{I}{}_{\mu}\rightarrow e'^{I}{}_{\mu}=\Lambda^{I}{}_{J}e^{J}{}_{\mu}$, those covariant derivatives $\mathcal{D}_{\mu}e^{I}{}_{\nu}$ are transformed as follows:
\begin{equation}
    \mathcal{D}_{\mu}e'^{I}{}_{\nu}=\Lambda^{I}{}_{J}\mathcal{D}_{\mu}e^{J}{}_{\nu}\,.
\label{Covariant derivative of co-vielbein}
\end{equation}
In the same manner, $\mathcal{D}_{\mu}e_{I}{}^{\nu}$ transforms as follows:
\begin{equation}
    \mathcal{D}_{\mu}e'_{I}{}^{\nu}=(\Lambda^{-1})^{J}{}_{I}\mathcal{D}_{\mu}e_{J}{}^{\nu}\,.
\label{Covariant derivative of vielbein}
\end{equation}
This is nothing but the property of the covariant derivative of gauge fields in the sense of the conventional gauge theory such like Yang-Mills theory.

Finally, let us derive the Weitzenb\"{o}ch connection. A section, denote $X$, of the tangent bundle $T\mathcal{M}$ is, of course, expressed as $X=X^{\mu}\partial_{\mu}$. Pulling back $X^{\mu}$ to the internal space bundle by using the co-frame fields $e^{I}{}_{\mu}$, the components of a section on the internal space are obtained: $\bar{X}^{I}=e^{I}{}_{\mu}X^{\mu}$, or equivalently, $X^{\mu}=e^{\mu}{}_{I}\bar{X}^{I}$. Therefore, $X$ is equivalent to $\bar{X}=\bar{X}^{I}\xi_{I}$. The covariant derivative of $X$ and $\bar{X}$ become as follows:
\begin{equation}
    \nabla X=(\partial_{\nu}X^{\mu}+\tilde{\Gamma}^{\mu}_{\nu\rho}X^{\rho})\partial_{\mu}\otimes dx^{\nu}
\label{nabla X}
\end{equation}
and 
\begin{equation}
    \mathcal{D}\bar{X}=(e^{I}{}_{\nu}\partial_{\mu}X^{\nu}+X^{\nu}\partial_{\mu}e^{I}{}_{\nu}+\omega^{I}{}_{J\mu}e^{J}{}_{\nu}X^{\nu})\xi_{I}\otimes dx^{\mu}\,,
\label{D X dual}
\end{equation}
respectively. Using the pull back of $\partial_{\mu}$ by the frame fields $e_{I}{}^{\mu}$, $\xi_{I}=e_{I}{}^{\mu}\partial_{\mu}$, the left-hand side of the second formula becomes as follows:
\begin{equation}
    \mathcal{D}\bar{X}=(\partial_{\nu}X^{\mu}+X^{\rho}e_{I}{}^{\mu}\partial_{\nu}e^{I}{}_{\rho}+\omega^{I}{}_{J\nu}e_{I}{}^{\mu}e^{J}{}_{\rho}X^{\rho})\partial_{\mu}\otimes dx^{\nu}\,.
\label{D X}
\end{equation}
This formula is, of course, the same as the first one; the covariant derivatives $\nabla X$ and $\mathcal{D}\bar{X}$ are also equivalent. Therefore, the following relation is derived~\cite{Carroll:2004st}:
\begin{equation}
    \tilde{\Gamma}^{\rho}_{\mu\nu}=e_{I}{}^{\rho}\partial_{\mu}e^{I}{}_{\nu}+\omega^{I}{}_{J\mu}e_{I}{}^{\rho}e^{J}{}_{\nu}\,.
\label{Weitzenboch connection}
\end{equation}
This is the so-called Weitzenb\"{o}ch connection~\cite{Weitzenboch:1923}, which relates the connection 1-form to the affine connection via the (co-)frame fields and vice versa. Notice that Eq.~(\ref{Weitzenboch connection}) implies $\mathcal{D}_{\mu}e^{I}{}_{\nu}=0$, or equivalently, $\mathcal{D}_{\mu}e_{I}{}^{\nu}=0$. These properties are sometimes called vielbein {\it postulate} but, as shown in the above, always satisfied. That is, it is not a {\it postulate}. Notice also that the covariant derivative $\nabla$ in Eq.~(\ref{nabla X}) and $\mathcal{D}$ in Eq.~(\ref{D X}) are equivalent, and, therefore, $\nabla$ satisfies the same rule as Eq.~(\ref{Covariant derivative of vielbein}). Namely, a theory which is composed of using the covariant derivative $\nabla$ with respect to the Weitzenb\"{o}ch connection Eq.~(\ref{Weitzenboch connection}) is a gauge invariant theory under the (co-)frame transformation as discussed in Eq.~(\ref{Gauge transformation of spin connection}).

\subsection{\label{03:02}A physical application of the gauge approach to gravity}
For the sake of providing the validity of this formulation in physics, let us give an application to teleparallel gravity as an example. In Refs.~\cite{Blixt:2018znp,Pati:2022nwi}, the authors proposed a novel formulation for introducing the teleparallel connection whose curvature vanishes. This property plays a crucial role in establishing the theories of teleparallel gravity. Before their works appear, there are two ways to realize this property; Vanishing the spin connection so-called Weitzenb\"{o}ch gauge condition; Imposing this property by using Lagrange multipliers. (See Refs in Ref.~\cite{Blixt:2018znp}.) They approached this realization by generalizing the first way: a non-vanishing spin connection is assumed. This possibility is engraved in Eq.~(\ref{Gauge transformation of spin connection}). Namely, the Weiztenb\"{o}ch gauge condition, $\omega^{I}{}_{J\mu}:=0$, can be transformed into $\omega'^{I}{}_{J\mu}=(\Lambda^{-1})^{I}{}_{K}\partial_{\mu}\Lambda^{K}{}_{J}$ by utilizing the co-frame transformation: $e^{I}{}_{\mu}\rightarrow e'^{I}{}_{\mu}=\Lambda^{I}{}_{J}e^{J}{}_{\mu}$. This indicates that the Lagrangian of the theory contains the group element (components): $\Lambda^{I}{}_{J}\in ISO(3\,,1)$. Therefore, at first glance, the theory obtains new configuration variables. The authors, however, showed that the field redefinition so-called Weitzenb\"{o}ch tetrad, which coincides with the inverse co-frame transformation in the current article, prevents this change of the configuration of the theory by calculating explicitly the canonical momentum variables with respect to $\Lambda^{I}{}_{J}$. Therefore, the spin connection is still a pure gauge and can be set to zero.

This assertion can also be verified without any explicit calculation when considering the gauge approach to gravity introduced in the previous section. Namely, any vector bundle has the so-called standard flat connection in a local region~\cite{Baez:1995sj}. This allows to exist the vanishing spin connection in the local region. Then, co-frame transformation introduces $\Lambda^{I}{}_{J}$ but, taking Eq.~(\ref{Covariant derivative of co-vielbein}) into account, it is obvious that the theory does not change in the same sense as the conventional gauge theories such as Yang-Mills theory. In this way, based on the gauge approach to gravity, a great amount of knowledge in the conventional gauge theories would be applied to the research on theories of metric-affine gravity. In particular, simplifying calculations would contribute to proceeding with the Hamiltonian analysis of these theories. 

\section{\label{04}Riemann-Cartan geometry: a unified-description of curvature and torsion 2-forms with the M\"{o}bius representation}
The curvature, torsion, and non-metricity 2-forms are defined in the internal space $\mathcal{V}$ thanks to the local property: $\mathcal{V}|_{U}\simeq T\mathcal{M}|_{U}$. (The conventional intruduction of these geometric quantities are reviewed in~\ref{02}.) Using Eq.~(\ref{Metric-Vielbein relation}) and Eq.~(\ref{Weitzenboch connection}) to replace the metric in the internal space $\mathcal{V}$ and the spin connection by that in the manifold $\mathcal{M}$ and the Weitzenb\"{o}ch connection, respectively, and contracting by Eq.~(\ref{Veilbein contranction}) all remaining the internal space indices, all these geometric quantities are moved to those in the tangent bundle $T\mathcal{M}$. The converse is, of course, valid. In this section, based on this formulation, we focus on investigating these quantities in the internal space $\mathcal{V}$, and then let us show that the curvature and torsion 2-forms can be unified by using the so-called M\"{o}bius representation. Finally, the subclasses of the Riemann-Cartan geometry are classified. 

\subsection{\label{04:01}M\"{o}bius representation of affine group}
Let us first consider the case of the affine group: $A(n\,;\,\mathbb{R})=T(n\,;\,\mathbb{R})\rtimes GL(n\,;\,\mathbb{R})$. This group forms an internal space bundle introduced in Sec.~\ref{03}. The group multiplication of $A(n\,;\mathbb{R})$, {\it i\,.e\,.,} the semi-direct product, denote ``$\,\circ\,$'', is defined as follows:
\begin{equation}
    (s_{1}\,,t_{1})\circ(s_{2}\,,t_{2})=(s_{1}\cdot s_{2}\,,s_{1}\cdot t_{2}+t_{1})\,,
\label{semi-directproduct}
\end{equation}
where $s_{1}=s_{1}(p)\,,s_{2}=s_{2}(p)\in GL(n\,;\,\mathbb{R})$, $t_{1}=t_{1}(p)\,,t_{2}=t_{2}(p)\in T(n\,;\,\mathbb{R})$, $p\in\mathcal{M}$, and `` $\cdot$ '' is the group multiplication of $T(n\,;\,\mathbb{R})$ and $GL(n\,;\,\mathbb{R})$, {\it i\,.e\,.,} the matrix product. The Lie algebra of this group, denoted as $\mathfrak{a}(n\,;\,\mathbb{R})=\mathfrak{t}(n\,;\,\mathbb{R})\rtimes \mathfrak{gl}(n\,;\,\mathbb{R})$, is generically given as follows:
\begin{equation}
\begin{split}
    &[P_{I}\,,P_{J}]={C^{(A1)}}^{K}{}_{IJ}P_{K}+{C^{(A2)}}^{K}{}_{IJL}E^{L}{}_{K}\,,\\
    &[E^{I}{}_{J}\,,P_{K}]={C^{(A3)}}^{IL}{}_{JK}P_{L}+{C^{(A4)}}^{IL}{}_{JKM}E^{M}{}_{L}\,,\\
    &[E^{I}{}_{J}\,,E^{K}{}_{L}]={C^{(A5)}}^{IKM}{}_{JL}P_{M}+{C^{(A6)}}^{IKN}{}_{JLM}E^{M}{}_{N}\,,
\end{split}
\label{Affine algebra}
\end{equation}
where $P_{I}$ and $E^{I}{}_{J}$ are the generator of the Lie algebras $\mathfrak{t}(n\,;\,\mathbb{R})$ and $\mathfrak{gl}(n\,;\,\mathbb{R})$, respectively. ${C^{(A1)}}^{K}{}_{IJ}$, ${C^{(A2)}}^{K}{}_{IJL}$, $\cdots$, ${C^{(A6)}}^{IKN}{}_{JLM}$ are structure constants of the Lie algebra of the affine group. The affine group demands that ${C^{(A3)}}^{IL}{}_{JK}=\delta^{I}{}_{K}\delta^{L}{}_{J}$, ${C^{(A6)}}^{IKN}{}_{JLM}=\delta^{I}{}_{L}\delta^{K}{}_{M}\delta^{N}{}_{J}-\delta^{I}{}_{M}\delta^{K}{}_{J}\delta^{N}{}_{L}$, and the other remaining structure constants vanish~\cite{Lord:1978}. The affine group $A(n\,;\,\mathbb{R})$ can be regarded as a subgroup of $GL(n+1\,;\,\mathbb{R})$ as follows:
\begin{equation}
    A_{\rm Mobius}(n\,;\,\mathbb{R})=
        \Bigg\{
                \begin{bmatrix}
                    s(p) & t(p) \\
                    0 & 1
                \end{bmatrix}
            \Bigg|\,
                s(p)\in GL(n\,;\,\mathbb{R})\,,t(p)\in T(n\,;\,\mathbb{R})
        \Bigg\}\,.
\label{Mebius representation}
\end{equation}
This representation is called the M\"{o}bius representation~\cite{Kobayashi:1972,Mielke:1987,Hehl:1994ue}. Then the ordinary matrix product of the elements of $A_{\rm Mobius}(n\,;\,\mathbb{R})$ restores the semi-direct product defined by Eq.~(\ref{semi-directproduct}) in as the first row of the result of the matrix product. In addition, acting an element $\Lambda(p)\in A_{\rm Mobius}$ on ${}^{T}(x\,,1)$, where $x=x(p)$ is the coordinates of the point $p\in\mathcal{M}$, it is revealed that the first component of the result is nothing but the affine transformation of the point $p$ as follows: $\Lambda(p)\,{}^{T}(x\,,1)={}^{T}(s(p)x(p)+t(p)\,,1)$. Therefore, the action of the affine group $A(n\,;\,\mathbb{R})$ leaves the $n$-dimensional hyperplane $\mathbb{R}^{n}$ invariant. 

In this representation, the affine connection is given as follows~\cite{Kobayashi:1996}:
\begin{equation}
    \omega^{(A)}=
    \begin{bmatrix}
        \omega^{(E)} & \omega^{(T)} \\
        0 & 0
    \end{bmatrix}
    =\begin{bmatrix}
        \omega^{(E)\,I}{}_{J}\otimes E^{J}{}_{I} & \omega^{(T)\,I}{}\otimes P_{I} \\
        0 & 0
    \end{bmatrix}\,,
\label{Mebius connection}
\end{equation}
where $\omega^{(E)}$ and $\omega^{(T)}$ are a $\mathfrak{gl}(n\,;\,\mathbb{R})$-valued 1-form on $\mathcal{M}$ and a $\mathfrak{t}(n\,;\,\mathbb{R})$-valued 1-form on $\mathcal{M}$, respectively. Then, for a frame transformation $e^{I}\rightarrow e'^{I}=\Lambda^{I}{}_{J}e^{J}$, a $\mathfrak{a}(n\,;\,\mathbb{R})$-valued 1-form $\omega^{(A)}$ on $\mathcal{M}$ is transformed as follows:
\begin{equation}
    \omega^{(A)}\rightarrow\omega'^{(A)}=\Lambda^{-1}d\Lambda+\Lambda^{-1}\omega^{(A)}\Lambda\,,
\label{Gauge transformation of Mebius connection}
\end{equation}
where $\Lambda\in A_{\rm Mebius}(n\,;\,\mathbb{R})$, or equivalently, 
\begin{equation}
\begin{split}
    &\omega^{(E)} \rightarrow \omega'^{(E)}=s^{-1}ds+s^{-1}\omega^{(E)}s\,,\\
    &\omega^{(T)} \rightarrow \omega'^{(T)}=s^{-1}d_{\nabla}t+s^{-1}\omega^{(T)}\,,\quad d_{\nabla}t=dt+d\rho(\omega^{(E)})t
\end{split}
\label{Separated gauge transformation of Mebius connection}
\end{equation}
where $s\in GL(n\,;\,\mathbb{R})$ and $t\in T(n\,;\,\mathbb{R})$. The first transformation law in Eq.~(\ref{Separated gauge transformation of Mebius connection}) indicates that $\omega^{(E)}$ is nothing but the Ehresmann connection of the principal $GL(n\,;\,\mathbb{R})$-bundle. Therefore, $d_{\nabla}t=dt+d\rho(\omega^{(E)})t$ is actually the covariant exterior derivative of $t$ with respect to the connection $\omega^{(E)}$, where $d\rho$ is a representation of the Lie algebra $\mathfrak{gl}(n\,;\,\mathbb{R}$). 

\subsection{\label{04:02}Potential 1-form of curvature and torsion 2-form}
The curvature 2-form is given as follows~\cite{Kobayashi:1972,Mielke:1987,Hehl:1994ue}:
\begin{equation}
    \Omega^{(A)}=d\omega^{(A)}+\omega^{(A)}\wedge\,\omega^{(A)}
    =\begin{bmatrix}
        \Omega^{(E)} & \Omega^{(T)}\,\\
        0 & 0 \,
    \end{bmatrix}
    =\begin{bmatrix}
        d\omega^{(E)}+\omega^{(E)}\wedge \omega^{(E)} & d\omega^{(T)}+\omega^{(E)}\wedge \omega^{(T)}\,\\
        0 & 0
    \end{bmatrix}\,.
\label{Mebius curvature}
\end{equation}
Then $\Omega^{(A)}$ is a $\mathfrak{gl}(n\,;\,\mathbb{R})$-valued 2-form on $\mathcal{M}$. For a frame transformation $e^{I}\rightarrow e'^{I}=\Lambda^{I}{}_{J}e^{J}$, $\Omega^{(A)}$ is transformed as follows:
\begin{equation}
    \Omega^{(A)} \rightarrow \Omega'^{(A)}=\Lambda^{-1}\Omega^{(A)}\Lambda\,.
\label{Pseudo gauge transformation of Mebius curvature}
\end{equation}
Remark that this is not a gauge transformation law.~\footnote{$\omega^{(T)}$ is not an Ehresmann connection of the principal $T(n\,;\,\mathbb{R})$-bundle. That is, only the form coincides with Eq.~(\ref{Gauge transformation of curvature 2-form}).}

Cartan's structure equation (See Eq.~(\ref{Torsion in local representation}), or equivalently, Eq.~(\ref{Torsion in component-form 2}) in detail) implies the meaning of $\Omega^{(T)}=d\omega^{(T)}+\omega^{(E)}\wedge \omega^{(T)}$. That is, if the $\omega^{(T)}$ is equivalent to the coframe field $\mathbf{e}^{-1}$, $\Omega^{(T)}$ is nothing but the torsion of the spacetime manifold $\mathcal{M}$. In order to confirm this conjecture, let us introduce an auxiliary vector field $\zeta=\zeta^{I}P_{I}\,$. Then the coframe field $\mathbf{e}^{-1}$ can be expressed as follows~\cite{Pilch:1979bi,Hehl:1994ue}:
\begin{equation}
    \mathbf{e}^{-1}=\omega^{(T)}+d_{\nabla}\zeta\,,
\label{Relation between Coframe and Translation}
\end{equation}
where $\mathbf{e}^{-1}=e^{I}\otimes P_{I}=e^{I}{}_{\mu}dx^{\mu}\otimes P_{I}$ and $\omega^{(T)}=\omega^{(T)\,I}{}_{}\otimes P_{I}=\omega^{(T)\,I}{}_{\mu}dx^{\mu}\otimes P_{I}$. Therefore, if the following condition holds then the statement is shown:
\begin{equation}
    d_{\nabla}\zeta:=0\,.
\label{Condition for Cartan connetion}
\end{equation}
Under the imposition of the above conditoin, the affine connection~(\ref{Mebius connection}) and the curvature 2-form become as follows~\cite{Kobayashi:1972}:
\begin{equation}
    \omega^{(C)}=
    \begin{bmatrix}
        \omega^{(E)} & \mathbf{e}^{-1} \\
        0 & 0
    \end{bmatrix}\,,\quad
    \Omega^{(C)}=d\omega^{(C)}+\omega^{(C)}\wedge \omega^{(C)}
    =\begin{bmatrix}
        \Omega^{(E)} & T\,\\
        0 & 0
    \end{bmatrix}\,,
\label{Cartan connection and curvature}
\end{equation}
where $\Omega^{(E)}$ and $T$ are given as follows:
\begin{equation}
    T=d\mathbf{e}^{-1}+\omega^{(E)}\wedge \mathbf{e}^{-1}=d_{\nabla}\mathbf{e}^{-1}\,,\quad \Omega^{(E)}=d\omega^{(E)}+\omega^{(E)}\wedge \omega^{(E)}=d_{\nabla}\omega^{(E)}\,.
\label{Riemann and Cartan geometry}
\end{equation}
These two equations are nothing but Cartan's first and second structure equations. This connection $\omega^{(C)}$ is the so-called Cartan connection~\cite{Cartan:1951,Kobayashi:1972}, although only the part $\omega^{(E)}$ in $\omega^{(C)}$ is the Ehresmann connection; the remaining part $\mathbf{e}^{-1}$ has nothing to do with Ehresmann connections, therefore, strictly speaking, this use of the word `connection' is just an abuse of jargon. Note that, for a frame transformation $e^{I}\rightarrow e'^{I}=\Lambda^{I}{}_{J}e^{J}$, $\Omega^{(E)}$ and $T$ transform as follows:
\begin{equation}
    \Omega^{(E)}\rightarrow \Omega'^{(E)}=\Lambda^{-1}\Omega^{(E)}\Lambda\,,\quad T\rightarrow T'=\Lambda^{-1}T\,,
\label{}
\end{equation}
respectively. Remark also that this decomposition is valid only in the satisfaction of Eq.~(\ref{Condition for Cartan connetion}). 

\subsection{\label{04:03}Classification of geometry}
The geometry based on the affine group $A(n\,;\,\mathbb{R})=T(n\,;\,\mathbb{R})\rtimes GL(n\,;\,\mathbb{R})$ is, therefore, the so-called (i) Riemann-Cartan geometry, which includes not only the curvature 2-form but also the torsion 2-form. The Riemann-Cartan geometry is denoted by ``$U_{n}$'', where $n$ is the dimension of the geometry. $U_{n}$ has four subclasses depending on whether or not the curvature 2-form and/or torsion 2-form vanishes; (ii-a) ``$T_{n}$'': The curvature 2-form vanishes: $\Omega^{(E)}:=0$ then the geometry turns into a teleparallel geometry; (iii) ``$V_{n}$'': The torsion 2-form vanishes: $T:=0$ then the geometry turns into a Riemann geometry; (iv) ``$E_{n}$'': The curvature and torsion 2-forms vanish: $\Omega^{(E)}:=0$ and $T:=0$ then the geometry turns into a Euclidean geometry. If the signature of the metric of the geometry is Lorentzian, ``$E_{n}$'' becomes a Minkowski geometry: ``$M_{n}$''. Therefore, the Riemann-Cartan geometry provides geometrical extensions of that of general relativity. In particular, the existence of torsion gives rise to a new subclass, {\it i\,.e\,.,} $T_{n}$, for theories of gravity. 

\section{\label{05}Extended M\"{o}bius representation 1: the unification of dilation 2-form, and its potential 1-form}
In this section, the M\"{o}bius representation of Weyl group is introduced, and the potential 1-form of the dilation 2-form is derived. The dilation 2-form provides the trace of the non-metricity 2-form. Finally, the subclasses of the Weyl geometry are classified. At least to the author's knowledge, no one has asserted this sort of approach to introducing dilation 2-form based on the extension of the M\"{o}bi\"{u}s representation of the affine group.

\subsection{\label{05:01}M\"{o}bius representation of Weyl group and potential 1-form of dilation 2-form}
Let us extend the affine group into the so-called Weyl group: $W(n\,;\,\mathbb{R})=D(n\,;\,\mathbb{R})\rtimes A(n\,;\,\mathbb{R})$. The Lie algebra of $W(n\,;\,\mathbb{R})$, {\it i\,.e\,.,} $\mathfrak{w}(n\,;\,\mathbb{R})=\mathfrak{d}(n\,;\,\mathbb{R})\rtimes \mathfrak{a}(n\,;\,\mathbb{R})$, is identified by the affine algebra~(\ref{Affine algebra}) together with the following generic algebra:
\begin{equation}
\begin{split}
    &[D\,,D]={C^{(W1)}}^{I}{}_{}P_{I}+{C^{(W2)}}^{I}{}_{J}E^{I}{}_{J}+C^{(W3)}D\,,\\ 
    &[D\,,P_{I}]={C^{(W4)}}^{J}{}_{I}P_{J}+{C^{(W5)}}^{J}{}_{IK}E^{K}{}_{J}+{C^{(W6)}}_{I}D\,,\\
    &[D\,,E^{I}{}_{J}]={C^{(W7)}}^{IK}{}_{J}P_{K}+{C^{(W8)}}^{IK}{}_{JL}E^{L}{}_{K}+{C^{(W9)}}^{I}{}_{J}D\,,    
\end{split}
\label{Special algebra of Weyl group}
\end{equation}
where $D$ is the generator of the Lie algebra of $D(n\,;\,\mathbb{R})$ and ${C^{(W1)}}^{I}{}_{}$, ${C^{(W2)}}^{I}{}_{J}$, $\cdots$, ${C^{(W9)}}$ are structure constants of the Lie algebra of the Weyl group. Herein, of course, ${C^{(W1)}}^{I}{}_{}$, ${C^{(W2)}}^{I}{}_{J}$, and $C^{(W3)}$ identically vanish. The Weyl algebra demands that ${C^{(W4)}}^{J}{}_{I}=\delta^{J}{}_{I}$ and the other remaining structure constants vanish.~\cite{Weyl:1918,Charap:1973fi,Hehl:1986ei} The M\"{o}bius representation of the Weyl group is given as follows:
\begin{equation}
    W_{\rm Mobius}(n\,;\,\mathbb{R})=
        \Bigg\{
                \begin{bmatrix}
                    s(p) & t(p) & \,\,\,{\bar\!\!\!d}(p) \\
                    0 & 1 & 0 \\
                    0 & 0 & 1
                \end{bmatrix}
            \Bigg|\,
                s(p)\in GL(n\,;\,\mathbb{R})\,,t(p)\in T(n\,;\,\mathbb{R})\,,\,\,\,{\bar\!\!\!d}\in D(n\,;\,\mathbb{R})
        \Bigg\}\,.
\label{Extended Mebius representation for Weyl}
\end{equation}
This is of course a subgroup of $GL(n+2\,;\,\mathbb{R})$. Notice that this representation introduces an extension of the semi-direct product Eq.~(\ref{semi-directproduct}) of the affine group into the Weyl group as follows:
\begin{equation}
    (s_{1}\,,t_{1}\,,\,\,\,{\bar\!\!\!d}_{1})\circ(s_{2}\,,t_{2}\,,\,\,\,{\bar\!\!\!d}_{2})=(s_{1}\cdot s_{2}\,,s_{1}\cdot t_{2}+t_{1}\,,s_{1}\cdot \,\,\,{\bar\!\!\!d}_{2}+\,\,\,{\bar\!\!\!d}_{1})\,.
\label{extended semi-directproduct for Weyl}
\end{equation}
Then, an extension of the Cartan connection~(\ref{Cartan connection and curvature}), let us call it the ``{\it Weyl connection}'', is defined as follows:
\begin{equation}
    \omega^{(W)}=
    \begin{bmatrix}
        \omega^{(E)} & \omega^{(T)} & \omega^{(D)}\\
        0 & 0 & 0 \\
        0 & 0 & 0
    \end{bmatrix}
    =\begin{bmatrix}
        \omega^{(E)\,I}{}_{J}\otimes E^{J}{}_{I} & \omega^{(T)\,I}{}\otimes P_{I} & \tilde{\omega}^{(D)}\otimes D\\
        0 & 0 & 0 \\
        0 & 0 & 0
    \end{bmatrix}\,,
\label{Extended Mebius connection for Weyl}
\end{equation}
where 
\begin{equation}
    \tilde{\omega}^{(D)}=\frac{1}{n}\,{\rm ln}(\left|{\rm det}(g)\right|)\,\mathbf{e}^{-1}
    \,.
\label{Dilation potential}
\end{equation}

The same consideration as the case of the affine group leads to the following transformation laws:
\begin{equation}
\begin{split}
    &\omega^{(E)} \rightarrow \omega'^{(E)}=s^{-1}ds+s^{-1}\omega^{(E)}s\,,\\
    &\omega^{(T)} \rightarrow \omega'^{(T)}=s^{-1}d_{\nabla}t+s^{-1}\omega^{(T)}\,,\quad d_{\nabla}t=dt+d\rho(\omega^{(E)})t\,,\\
    &\omega^{(D)} \rightarrow \omega'^{(D)}=s^{-1}d_{\nabla}\,\,\,{\bar\!\!\!d}+s^{-1}\omega^{(D)}\,,\quad d_{\nabla}\,\,\,{\bar\!\!\!d}=d\,\,\,{\bar\!\!\!d}+d\rho(\omega^{(D)})\,\,\,{\bar\!\!\!d}
\end{split}
\label{Separated gauge transformation of Weyl connection}
\end{equation}
for the frame transformation $e^{I}\rightarrow e'^{I}=\Lambda^{I}{}_{J}e^{J}$. Therefore, the $\mathfrak{d}(n\,;\,\mathbb{R})$-valued 1-form $\omega^{(D)}$ is not an Ehresmann connection. Alternatively, this valued 1-form becomes a potential to generate the following geometric quantity:
\begin{equation}
    \Delta=d_{\nabla}\omega^{(D)}-\frac{1}{n}\,{\rm ln}(\left|{\rm det}(g)\right|)\,T=\frac{1}{n}\,{\rm Tr}\,(\,d_{\nabla}\,g\,)\,\mathbf{e}^{-1}=\frac{1}{n}\,{\rm Tr}\,(\,Q\,)\,\mathbf{e}^{-1}\,,
\label{Dilation}
\end{equation}
where ``$\,\rm Tr\,$'' is the trace operator and $Q=d_{\nabla}g$ is the non-metricity 2-form. (See Eq.~(\ref{Non-metricity tensor from potential}) in detail.) Where we abbreviated `` $\,\tilde{}\,$ '' of each geometric quantity. Remark that the commutativity of the generators $D$ and $E^{I}{}_{J}$ ensures $\omega^{(E)}\wedge \omega^{(D)}=0$. This quantity $\Delta$ is the so-called dilation 2-form. Notice that if the torsion 2-form vanishes: $T=0$, Eq.~(\ref{Dilation}) turn to simply to be
\begin{equation}
    \Delta=d_{\nabla}\omega^{(D)}=\frac{1}{n}\,{\rm Tr}\,(\,d_{\nabla}\,g\,)\,\mathbf{e}^{-1}=\frac{1}{n}\,{\rm Tr}\,(\,Q\,)\,\mathbf{e}^{-1}\,.
\label{Dilation under torsion-free}
\end{equation}
The torsion-free condition can be realized by taking the auxiliary field $\zeta$ in Eq.~(\ref{Relation between Coframe and Translation}) as follows:
\begin{equation}
    d_{\nabla}\omega^{(T)}+{d_{\nabla}}^{2}\zeta:=0\,.
\label{Condition for Weyl geometry}
\end{equation}
Remark that ${d_{\nabla}}^{2}=d_{\nabla}d_{\nabla}$ does not vanish unlike the ordinary exterior derivative: $d^{2}=dd=0$. Based on the classification in Sec.~\ref{04:03}, such geometry can contain the subclasses $V_{n}$ and $E_{n}$ (or $M_{n}$). 

\subsection{\label{05:02}Classification of geometry}
The geometry based on the Weyl group $W(n\,;\,\mathbb{R})=D(n\,;\,\mathbb{R})\rtimes A(n\,;\,\mathbb{R})$ is richer than the Riemann-Cartan geometry by virtue of the existence of the dilation 2-form. This geometry is called (v) Weyl-Cartan geometry and denoted as ``$Y_{n}$''. On one hand, $Y_{n}$ contains $U_{n}$ as a special case of vanishing dilation 2-form: $\Delta:=0$. On the other hand, $Y_{n}$ gives new subclasses; (vi-a) ``$W_{n}$'': The torsion 2-form vanishes: $T:=0$ then the geometry turns into a Weyl geometry; (ii-b) $T_{n}$: The curvature 2-form vanishes: $\Omega^{(E)}:=0$ then the geometry turns into one of the teleparallel geometry together with the non-vanishing dilation 2-form. Therefore, the Weyl-Cartan geometry provides geometrical extensions of that of general relativity. In particular, the subclass $W_{n}$, on one hand, provides the geometrical extension but generically associates the dilation 2-form. On the other hand, the subclass $T_{n}$ departs geometrically from that of general relativity. Finally, notice that the dilation of the Weyl geometry, $W_{n}$, is given by Eq.~(\ref{Dilation under torsion-free}). 

\section{\label{06}Extended M\"{o}bius representation 2: the unification of non-metricity 2-form, and its potential 1-form}
In this section, the M\"{o}bius representation of metric-affine group is introduced, and the potential 1-form of the shear 2-form is derived. The shear 2-form provides the non-metricity 2-form together with the dilation 2-form. Finally, the subclasses of the metric-affine geometry are classified. At least to the author's knowledge, no one has asserted this sort of approach to introducing dilation and shear 2-form based on the extension of the M\"{o}bi\"{u}s representation of the Weyl group.

\subsection{\label{06:01}M\"{o}bius representation of metric-affine group and potential 1-form of non-metricity 2-form}
Further generalization of the Weyl geometry is possible; the dilation 2-form given in Eq.~(\ref{Dilation}) implies the existence of a geometry such that the non-metricity 2-form is generated from an exterior covariant derivative of some quantity of potential. Let us consider the metric-affine group: $MA(n\,;\,\mathbb{R})=S(n\,;\,\mathbb{R})\rtimes W(n\,;\,\mathbb{R})$. The Lie algebra of $MA(n\,;\,\mathbb{R})$, $\mathfrak{ma}(n\,;\,\mathbb{R})=\mathfrak{s}(n\,;\,\mathbb{R})\rtimes \mathfrak{w}(n\,;\,\mathbb{R})$, is identified by the affine algebra~(\ref{Affine algebra}), the Weyl algebra~(\ref{Special algebra of Weyl group}), and the following generic algebra:
\begin{equation}
\begin{split}
    &[S^{I}{}_{J}\,,P_{K}]={C^{(MA1)}}^{IL}{}_{JK}P_{L}+{C^{(MA2)}}^{IL}{}_{JKM}E^{M}{}_{L}+{C^{(MA3)}}^{I}{}_{JK}D+{C^{(MA4)}}^{IL}{}_{JKM}S^{M}{}_{L}\,,\\
    &[S^{I}{}_{J}\,,E^{K}{}_{L}]={C^{(MA5)}}^{IKM}{}_{JL}P_{M}+{C^{(MA6)}}^{IKM}{}_{JLN}E^{N}{}_{M}+{C^{(MA7)}}^{IK}{}_{JL}D+{C^{(MA8)}}^{IKM}{}_{JLN}S^{N}{}_{M}\,,\\
    &[S^{I}{}_{J}\,,D]={C^{(MA9)}}^{IK}{}_{J}P_{K}+{C^{(MA10)}}^{IK}{}_{JL}E^{L}{}_{K}+{C^{(MA11)}}^{I}{}_{J}D+{C^{(MA12)}}^{IK}{}_{JL}S^{L}{}_{K}\,,\\
    &[S^{I}{}_{J}\,,S^{K}{}_{L}]={C^{(MA13)}}^{IKM}{}_{JL}P_{M}+{C^{(MA14)}}^{IKM}{}_{JLN}E^{N}{}_{M}+{C^{(MA15)}}^{IK}{}_{JL}D+{C^{(MA16)}}^{IKM}{}_{JLN}S^{N}{}_{M}\,,    
\end{split}
\label{Special algebra of Metric-affine group}
\end{equation}
where $S^{I}{}_{J}$ are the generator of the Lie algebra of $S(n\,;\,\mathbb{R})$, let us call it the ``{\it shear group}'', and ${C^{(MA1)}}^{IL}{}_{JK}$, ${C^{(MA2)}}^{IKM}{}_{JLN}$, $\cdots$, ${C^{(MA16)}}^{IKM}{}_{JLN}$ are structure constants of the Lie algebra of the metric-affin group. The metric-affine algebra demands that ${C^{(MA1)}}^{IL}{}_{JK}=\delta^{I}{}_{K}\delta^{L}{}_{J}$ and the other remaining structure constants vanish. Notice that this algebra is nothing but a generalization of the Weyl algebra~(\ref{Special algebra of Weyl group}); in the case of its dimension being one, Eq.~(\ref{Special algebra of Metric-affine group}) results in Eq.~(\ref{Special algebra of Weyl group}). Then the M\"{o}bius representation of the metric-affine group is given as follows:
\begin{equation}
\begin{split}
    {MA}_{\rm \,\,Mobius}(n\,;\,\mathbb{R})=
        \Bigg\{
                \begin{bmatrix}
                    s(p) & t(p) & \,\,\,{\bar\!\!\!d}(p) & \,\,\,{\shortmid\!\!\!\!s}(p) \\
                    0 & 1 & 0 & 0 \\
                    0 & 0 & 1 & 0 \\
                    0 & 0 & 0 & 1
                \end{bmatrix}
            \Bigg|\,&
                s(p)\in GL(n\,;\,\mathbb{R})\,,t(p)\in T(n\,;\,\mathbb{R})\,,\\
                &\,\,\,{\bar\!\!\!d}\in D(n\,;\,\mathbb{R})\,,\,\,\,{\shortmid\!\!\!\!s}\in S(n\,;\,\mathbb{R})
        \Bigg\}\,.    
\end{split}
\label{Extended Mebius representation for metric-affine}
\end{equation}
This is a subgroup of $GL(2(n+1)\,;\,\mathbb{R})$. Notice that this representation introduces an extension of the extended semi-direct product Eq.~(\ref{extended semi-directproduct for Weyl}) of the affine group into the metric-affine group as follows:
\begin{equation}
    (s_{1}\,,t_{1}\,,\,\,\,{\bar\!\!\!d}_{1}\,,\,\,\,{\shortmid\!\!\!\!s}_{1})\circ(s_{2}\,,t_{2}\,,\,\,\,{\bar\!\!\!d}_{2}\,,\,\,\,{\shortmid\!\!\!\!s}_{2})=(s_{1}\cdot s_{2}\,,s_{1}\cdot t_{2}+t_{1}\,,s_{1}\cdot \,\,\,{\bar\!\!\!d}_{2}+\,\,\,{\bar\!\!\!d}_{1}\,,s_{1}\cdot\,\,\,{\shortmid\!\!\!\!s}_{2}+\,\,\,{\shortmid\!\!\!\!s}_{1})\,.
\label{extended semi-directproduct for metric-affine}
\end{equation}
Then, an extension of the Weyl connection, let us call it the ``{\it metric-affine connection}'', is introduced as follows:
\begin{equation}
    \omega^{(MA)}=
    \begin{bmatrix}
        \omega^{(E)} & \omega^{(T)} & \omega^{(D)} & \omega^{(S)} \\
        0 & 0 & 0 & 0 \\
        0 & 0 & 0 & 0 \\
        0 & 0 & 0 & 0
    \end{bmatrix}
    =\begin{bmatrix}
        \omega^{(E)\,I}{}_{J}\otimes E^{J}{}_{I} & \omega^{(T)\,I}{}\otimes P_{I} & \omega^{(D)}\otimes D & \omega^{(S)I}{}_{J}\otimes S^{J}{}_{I} \\
        0 & 0 & 0 & 0 \\
        0 & 0 & 0 & 0 \\
        0 & 0 & 0 & 0
    \end{bmatrix}\,,
\label{Extended Mebius connection for metric-affine}
\end{equation}
where 
\begin{equation}
\begin{split}
    &\omega^{(S)I}{}_{J}=\tilde{\omega}^{(S)}\delta^{I}{}_{J}\,,\\
    &\tilde{\omega}^{(S)}=\left(g-\frac{1}{n}{\rm ln}(\left|{\rm det}(g)\right|)\right)\mathbf{e}^{-1}
    \,.
\end{split}
\label{Shear potential}
\end{equation}
For a frame transformation, $\omega^{(S)}$ obeys the same rule as $\omega^{(T)}$, $\omega^{(D)}$ in Eq.~(\ref{Separated gauge transformation of Weyl connection}). Then the $\mathfrak{s}(n\,;\,\mathbb{R})$-valued 1-form $\omega^{(S)}$ becomes a potential to generate the following geometric quantity:
\begin{equation}
    {\nearrow\!\!\!\!\!\!\!\Delta}=d_{\nabla}\omega^{(S)}-\left(g-\frac{1}{n}{\rm ln}(\left|{\rm det}(g)\right|)\right)T=Q\,\mathbf{e}^{-1}-\Delta=\left(Q-\frac{1}{n}\,{\rm Tr}\,(\,Q\,)\right)\,\mathbf{e}^{-1}\,.
\label{Shear}
\end{equation}
This quantity ${\nearrow\!\!\!\!\!\!\!\Delta}$ is the so-called shear 2-form. Where we abbreviated `` $\,\tilde{}\,$ '' of each geometric quantity. Remark that the commutativity of the generators $D$ and $E^{I}{}_{J}$ ensures $\omega^{(E)}\wedge \omega^{(S)}=0$. Notice that the case of vanishing torsion turns Eq.~(\ref{Shear}) to be 
\begin{equation}
    {\nearrow\!\!\!\!\!\!\!\Delta}=d_{\nabla}\omega^{(S)}=Q\,\mathbf{e}^{-1}-\Delta=\left(Q-\frac{1}{n}\,{\rm Tr}\,(\,Q\,)\right)\,\mathbf{e}^{-1}\,.
\label{Shear under torsion-free}
\end{equation}
Taking into account the dilation 2-form, the non-metricity 2-form is restored as follows;
\begin{equation}
    Q={\nearrow\!\!\!\!\!\!\!Q}+\frac{1}{n}{\rm Tr}\,(Q)\,
\label{non-metricity}
\end{equation}
as a 1-form on $\mathcal{M}$, where ${\nearrow\!\!\!\!\!\!\!Q}$ is defined as ${\nearrow\!\!\!\!\!\!\!\Delta}={\nearrow\!\!\!\!\!\!\!Q}\,\mathbf{e}^{-1}$ under Eq.~(\ref{Shear}). Therefore, the existence of both the dilation and shear 2-forms is equivalent to that of the non-metricity 2-form, and vice versa. Notice, finally, that Eq.~(\ref{non-metricity}) holds not depending on whether or not Condition~(\ref{Condition for Weyl geometry}) is satisfied. 

\subsection{\label{06:02}Classification of geometry}
The geometry based on the metric-affine group $MA(n\,;\,\mathbb{R})=S(n\,;\,\mathbb{R})\rtimes W(n\,;\,\mathbb{R})$ is the most generic geometry by virtue of the existence not only the dilation 2-form but also the shear 2-form. This geometry is nothing but (vii) Metric-affine geometry and denoted as ``$L_{n}$''. On one hand, $L_{n}$ contains $Y_{n}$ as a special case of vanishing  shear 2-form: ${\nearrow\!\!\!\!\!\!\!\Delta}:=0$. On the other hand, $L_{n}$ gives new subclasses: (vi-b) $W_{n}$: The torsion 2-form vanishes: $T:=0$ then the geometry turns into the most generic subclass of the Weyl geometry together with the non-vanishing shear 2-form; (viii) ``$S_{n}$'': The torsion and curvature 2-forms vanish: $T:=0$ and $\Omega^{(E)}:=0$ then the geometry turns into the symmetric teleparallel one; (ii-c) $T_{n}$: The curvature 2-form vanishes: $\Omega^{(E)}:=0$ then the geometry turns into the most generic subclass of the teleparallel geometry together with the non-vanishing dilation and shear 2-forms: the non-metricity 2-form. Therefore, the metric-affine geometry provides the widest class of geometry in terms of curvature, torsion, and non-metricity 2-forms, and it contains that of general relativity as a special case. In particular, the subclass $S_{n}$ gives a geometric departure from general relativity. 

\subsection{\label{06:03}A physical application of the unified-description to gravity}
For providing the validity of this formulation in physics, let us illustrate briefly an important application to teleparallel gravity. As mentioned in Sec.~\ref{03:02}, teleparallel gravity demands the Weitzenb\"{o}ch gauge condition to realise the vanishing curvature. The unified-description of curvature and torsion given in Sec.~\ref{04} suggests that the torsion survives due to the first term of the first formula in Eq.~(\ref{Riemann and Cartan geometry}) even if the gauge realises the condition of the vanishing curvature: $T=d\mathbf{e}^{-1}$, or in component form, $T^{I}{}_{\mu\nu}=2\partial_{[\mu}e^{I}{}_{\nu]}$. TEGR (See Sec.~\ref{01}) uses this non-vanishing torsion to establish the theory of gravity. However, notice that this unified-description of curvature and torsion only is not enough to consider non-metricity. That was the motivation on extending this unified-description. As shown in Sec.~\ref{05} and Sec.~\ref{06}, extending the M\"{o}bi\"{u}s representation, dilation and shear (, or equivalently, non-metricity) can be taken into account. Then we consider another counterpart, STEGR (See Sec.~\ref{01}), by imposing the vanishing torsion condition: $e^{I}{}_{\mu}=\partial_{\mu}\eta^{I}$, where $\eta^{I}$ are St\"{u}kerberg fields. As will be shown in Sec.~\ref{07:02}, on one hand, this formulation leads to the so-called coincident gauge~\cite{BeltranJimenez:2022azb}. On the other hand, in TEGR, thanks to this extended unified-description, new gauge conditions for introducing the vanishing non-metricity condition can be clarified. We will show in Sec.~\ref{07:02} that the conventional internal metric, the Minkowski metric, is not the only choice to realise this condition. This new perspective would gives new insight into formulating the theories of gravity in terms of torsion. 

\section{\label{07}In\"{o}n\"{u}-Wigner contraction of the metric-affine geometry and its extensions}
In this section, based on the unified-description of curvature, torsion, and non-metricity (or equivalently, dilation and shear), the relationships between the algebraic structure and geometric quantities of each geometry are scrutinized by using the so-called In\"{o}n\"{u}-Wigner contraction. The ultimate purpose is to beyond the metric-affine geometry and to pursue new class of geometry and its gauge theory of gravity. First, the In\"{o}n\"{u}-Wigner contraction is introduced. Second, the contractions of the metric-affine geometry and its extension are performed. Finally, the de Sitter/anti-de Sitter cases are discussed. 

\subsection{\label{07:01}In\"{o}n\"{u}-Wigner contraction}
In\"{o}n\"{u}-Wigner contraction is a method of parametrizing a Lie algebra to decompose it into subalgebras~\cite{Segal:1951,Wigner:1953,Lord:1985,Evelyn:1995,Khasanov:2011jr,Subag:2012}. The contraction is originally valid not only for Lie algebras but also for those representations, but, in this paper, we discuss only the contraction of Lie algebras.

Let $G$ be a Lie group and $\mathfrak{g}$ be its Lie algebra. Then an element $g=g(\tau_{i})\in G$, where $\tau_{i}$ $(i=1\,,2\,,\cdots\,,t)$ are the number of $t$ group parameters, leads to an element of the Lie algebra of $G$ as its derivative with respect to the unit element $e\in G$. Let $I_{i}$ be such element of the Lie algebra. Then the following commutation relations hold:
\begin{equation}
    [I_{i}\,,I_{j}]=C^{k}{}_{ij}I_{k}\,,
\label{Generic Lie algebra}
\end{equation}
where $[\,\cdot\,,\cdot\,]$ is the Lie bracket and $C^{k}{}_{ij}$ are the structure constants of the Lie algebra. For this algebra, a transformation of the elements $I_{i}$ and the group parameters $\tau_{i}$ is introduced as follows:
\begin{equation}
    I'_{i}=P^{j}{}_{i}I_{j}\,,\quad \tau'_{i}=P^{j}{}_{i}\tau_{j}\,,
\label{Wigner transformation}
\end{equation}
where $P^{j}{}_{i}$ is a parametrization matrix. If the matrix is non-singular then the above transformation is just an automorphism of the Lie algebra. However, if it is singular, the situation gets changed; a subalgebra can be obtained. In particular, it is an intriguing case that the singularity is realized as some limits of the parameters. 

A simple case is the contraction by the following matrix:
\begin{equation}
    P^{i}{}_{j}=
        \begin{bmatrix}
            1_{r\times r} & 0_{r\times (n-r)}\\
            0_{(n-r)\times r} & 0_{(n-r)\times (n-r)}
        \end{bmatrix}
        +
        \begin{bmatrix}
            \epsilon^{\alpha}{}_{\beta} & 0_{r\times (n-r)}\\
            0_{(n-r)\times r} & \epsilon^{\bar{\alpha}}{}_{\bar{\beta}}
        \end{bmatrix}\,,
\label{An example of Wigner transformation}
\end{equation}
where $\epsilon^{\alpha}{}_{\beta}\in (0\,,1]$, $\epsilon^{\bar{\alpha}}{}_{\bar{\beta}}\in (0\,,1]$, $\alpha\,,\beta\in\{1\,,2\,,\cdots\,,t'<t\}$, and $\bar{\alpha}\,,\bar{\beta}\in\{t'+1\,,t'+2\,,\cdots\,,t\}$. Then, on one hand, the elements $I_{i}$ and those algebras are transformed as follows:
\begin{equation}
    I'_{i}=(\delta^{\alpha}{}_{i}+\delta^{\beta}{}_{i}\epsilon^{\alpha}{}_{\beta})I_{\alpha}+\delta^{\bar{\beta}}{}_{i}\epsilon^{\bar{\alpha}}{}_{\bar{\beta}}I_{\bar{\alpha}}\,,
\label{Wigner transformation of generators}
\end{equation}
and 
\begin{equation}
    [I'_{i}\,,I'_{j}]_{\epsilon}=\delta^{\alpha}{}_{i}\delta^{\beta}{}_{i}C^{\gamma}{}_{\alpha\beta}I_{\gamma}+\mathcal{O}(\epsilon)\,,
\label{Wigner transformation of algebras of generators}
\end{equation}
respectively, where $[\,\cdot\,,\cdot\,]_{\epsilon}$ denotes the Lie brackets of the transformed elements $I'_{i}$. Let $\mathfrak{g}_{\epsilon}$ and $G_{\epsilon}$ denote a new Lie algebra that is generated by $I'_{i}$ with satisfying the above algebra and the Lie group of $\mathfrak{g}_{\epsilon}$, respectively. Notice that $\mathfrak{g}_{\epsilon}$ and $G_{\epsilon}$ include $\mathfrak{g}$ and $G$ as a special case of $\epsilon^{\alpha}{}_{\beta}\rightarrow+0$ and $\epsilon^{\bar{\alpha}}{}_{\bar{\beta}}\rightarrow\delta^{\bar{\alpha}}{}_{\bar{\beta}}$, where `` $+0$ '' means the right-limit to zero in $(0\,,1]$. Therefore, if the limit of the transformation of the above algebras with respect to $\epsilon^{\alpha}{}_{\beta}\rightarrow+0$ and $\epsilon^{\bar{\alpha}}{}_{\bar{\beta}}\rightarrow+0$ exists then the following new algebra is obtained:
\begin{equation}
    [I'_{\alpha}\,,I'_{\beta}]_{\epsilon\rightarrow+0}=C^{\gamma}{}_{\alpha\beta}I'_{\gamma}\,.
\label{Wigner constraction of Wigner transformation of algebras of generators}
\end{equation}
That is, a new subalgebra $\mathfrak{g}_{0}=\mathfrak{g}_{\epsilon\rightarrow+0}=\left<I'_{\bar{\alpha}}\right>$ is obtained, and $\mathfrak{g}_{0}$ generates the Lie group $G_{0}=G_{\epsilon\rightarrow+0}$. On the other hand, the group parameters $\tau_{i}$ are contracted as follows:
\begin{equation}
\begin{split}
    &\tau'_{\alpha}=\tau_{\alpha}+\epsilon^{\beta}{}_{\alpha}\tau_{\beta}\rightarrow\tau_{\alpha}\quad (\epsilon^{\beta}{}_{\alpha}\rightarrow+0)\,,\\
    &\tau'_{\bar{\alpha}}=\epsilon^{\bar{\beta}}{}_{\bar{\alpha}}\tau_{\bar{\beta}}\rightarrow0\quad (\epsilon^{\bar{\beta}}{}_{\bar{\alpha}}\rightarrow+0)\,.
\end{split}
\label{Wigner contraction of group parameters}
\end{equation}
Notice that the group parameters $\tau'_{\bar{\alpha}}$ of the group $G_{\epsilon}$ converge to zero while remaining the corresponding parameters $\tau_{\bar{\beta}}$ of $G$. Since $G_{\epsilon}$ includes not only $G$ but also $G_{0}$, this means that the parametrization given in Eq.~(\ref{Wigner transformation}) extended the original group $G$ into a larger group $G_{\epsilon}$ as a topological space. 

\subsection{\label{07:02}Contraction of metric-affine algebra: New gauge conditions on non-metricity and revisiting to Weitzenb\"{o}ch and coincident gauge}
The Lie algebra of $MA(n\,;\,\mathbb{R})$ is summarized as follows:
\begin{equation}
\begin{split}
    &[P_{I}\,,P_{J}]=0\,,\quad [E^{I}{}_{J}\,,P_{K}]=\delta^{I}{}_{K}P_{J}\,,\quad [E^{I}{}_{J}\,,E^{K}{}_{L}]=\delta^{I}{}_{L}E^{K}{}_{J}-\delta^{K}{}_{J}E^{I}{}_{L}\,,\\
    &[D\,,D]=0\,,\quad [D\,,P_{I}]=P_{I}\,,\quad [D\,,E^{I}{}_{J}]=0\,,\\
    &[S^{I}{}_{J}\,,P_{K}]=\delta^{I}{}_{K}P_{J}\,,\quad [S^{I}{}_{J}\,,E^{K}{}_{L}]=0\,,\quad [S^{I}{}_{J}\,,D]=0\,,\quad [S^{I}{}_{J}\,,S^{K}{}_{L}]=0\,.
\end{split}
\label{Lie algebra of metric-affine group}
\end{equation}
A parametrization for the In\"{o}n\"{u}-Wigner contraction of the above algebra can be set as follows:
\begin{equation}
    P'_{I}=\epsilon^{(P)}P_{I}\,,\quad E'^{I}{}_{J}=\epsilon^{(E)}E^{I}{}_{J}\,,\quad D'=\epsilon^{(D)}D\,,\quad S'^{I}{}_{J}=\epsilon^{(S)}S^{I}{}_{J}\,,
\label{Parametrization for IW contraction}
\end{equation}
where $\epsilon^{(P)}$, $\epsilon^{(E)}$, $\epsilon^{(D)}$, and $\epsilon^{(S)}$ are a set of parameters and belong to the range $(0\,,1]$. Hereinafter, let us denote the parameter space spanned by the above parameters as\\
$\epsilon_{\rm \ type\ of\ contraction/geometry\ in\ MA}=(\epsilon^{(P)}\,,\epsilon^{(E)}\,,\epsilon^{(D)}\,,\epsilon^{(S)})$, where ``{type of contraction/geometry in MA}'' denotes a geometry which is obtained by performing the contraction. Hereinafter, we use this notation. Then the above algebras are parameterized as follows:
\begin{equation}
\begin{split}
    &[P'_{I}\,,P'_{J}]=0\,,\quad [E'^{I}{}_{J}\,,P'_{K}]=\epsilon^{(E)}\delta^{I}{}_{K}P'_{J}\,,\quad [E'^{I}{}_{J}\,,E'^{K}{}_{L}]=\epsilon^{(E)}(\delta^{I}{}_{L}E'^{K}{}_{J}-\delta^{K}{}_{J}E'^{I}{}_{L})\,,\\
    &[D'\,,D']=0\,,\quad [D'\,,P'_{I}]=\epsilon^{(D)}P'_{I}\,,\quad [D'\,,E'^{I}{}_{J}]=0\,,\\
    &[S'^{I}{}_{J}\,,P'_{K}]=\epsilon^{(S)}\delta^{I}{}_{K}P'_{J}\,,\quad [S'^{I}{}_{J}\,,E'^{K}{}_{L}]=0\,,\quad [S'^{I}{}_{J}\,,D']=0\,,\quad [S'^{I}{}_{J}\,,S'^{K}{}_{L}]=0\,.
\end{split}
\label{Parametrized Lie algebra of metric-affine group}
\end{equation}
Notice that this algebra does not explicitly contain the parameter $\epsilon^{(P)}$. Therefore, it is enough to express the parameter space without $\epsilon^{(P)}$ as $\epsilon_{\rm \ type\ of\ contraction/geometry\ in\ MA}=(\epsilon^{(E)}\,,\epsilon^{(D)}\,,\epsilon^{(S)})$, and there are eight possible contractions.
Of course, if all parameters are taken to be unity then the above algebra~(\ref{Parametrized Lie algebra of metric-affine group}) results in the original algebra~(\ref{Lie algebra of metric-affine group}). In terms of the parameter space notation, the algebra is expressed as $\epsilon_{\rm \ L_{n}}=(1\,,1\,,1)$.

The contraction generated by the parameters $\epsilon_{\rm \ Y_{n}}=(1\,,1\,,+0)$ derives the following algebras:
\begin{equation}
\begin{split}
    &[P'_{I}\,,P'_{J}]=0\,,\quad [E'^{I}{}_{J}\,,P'_{K}]=\delta^{I}{}_{K}P'_{J}\,,\quad [E'^{I}{}_{J}\,,E'^{K}{}_{L}]=\delta^{I}{}_{L}E'^{K}{}_{J}-\delta^{K}{}_{J}E'^{I}{}_{L}\,,\\
    &[D'\,,D']=0\,,\quad [D'\,,P'_{I}]=P'_{I}\,,\quad [D'\,,E'^{I}{}_{J}]=0\,,\\
    &[S'^{I}{}_{J}\,,P'_{K}]=0\,,\quad [S'^{I}{}_{J}\,,E'^{K}{}_{L}]=0\,,\quad [S'^{I}{}_{J}\,,D']=0\,,\quad [S'^{I}{}_{J}\,,S'^{K}{}_{L}]=0\,.
\end{split}
\label{Lie algebra of Weyl group}
\end{equation}
This is nothing but the algebra of the Weyl geometry. In this algebra, the shear given in Eq.~(\ref{Shear}) vanishes, {\it i\,.e\,.,} ${\nearrow\!\!\!\!\!\!\!\Delta}=0$, for the metric tensor of the internal space with satisfying the following equations:
\begin{equation}
    g^{IK}\partial_{\mu}g_{KJ}=\frac{1}{n}\,\delta^{I}{}_{J}\,g^{LM}\partial_{\mu}g_{LM}\,.
\label{Gauge-fixing for vanishing shear}
\end{equation}
The contraction generated by the parameters $\epsilon_{\rm \ Y_{n}\ with\ vanishing\ dilation}=(1\,,+0\,,1)$ leads to the following algebras:
\begin{equation}
\begin{split}
    &[P'_{I}\,,P'_{J}]=0\,,\quad [E'^{I}{}_{J}\,,P'_{K}]=\delta^{I}{}_{K}P'_{J}\,,\quad [E'^{I}{}_{J}\,,E'^{K}{}_{L}]=\delta^{I}{}_{L}E'^{K}{}_{J}-\delta^{K}{}_{J}E'^{I}{}_{L}\,,\\
    &[D'\,,D']=0\,,\quad [D'\,,P'_{I}]=0\,,\quad [D'\,,E'^{I}{}_{J}]=0\,,\\
    &[S'^{I}{}_{J}\,,P'_{K}]=\delta^{I}{}_{K}P'_{J}\,,\quad [S'^{I}{}_{J}\,,E'^{K}{}_{L}]=0\,,\quad [S'^{I}{}_{J}\,,D']=0\,,\quad [S'^{I}{}_{J}\,,S'^{K}{}_{L}]=0\,.
\end{split}
\label{Lie algebra of metric-affine group without dilation}
\end{equation}
Then the dilation given in Eq.~(\ref{Dilation}) vanishes, {\it i\,.e\,.,} $\Delta=0$ for the metric tensor of the internal space with satisfying the following equations:
\begin{equation}
    g^{IJ}\partial_{\mu}g_{IJ}=0\,.
\label{Gauge-fixing for vanishing dilation}
\end{equation}
Therefore, choosing a constant metric tensor, $g_{IJ}=c_{IJ}$, not restricting to the Minkowski metric, is a possible gauge condition for vanishing non-metricity and let us call this gauge condition ``{\it trivial gauge}''. This property unveils also a new geometric description that the non-vanishing shear and/or dilation, or equivalently, non-metricity allows to choose $g_{IJ}\neq c_{IJ}$ as another possible gauge condition instead of the trivial gauge. Remark, here, that Conditions~(\ref{Gauge-fixing for vanishing shear}) and~(\ref{Gauge-fixing for vanishing dilation}) can be independently imposed. These statements will be investigated in detail in a sequel paper.

In the same manner, the contraction generated by $\epsilon_{\rm \ U_{n}}=(1\,,+0\,,+0)$ derives the algebra of the Riemann-Cartan geometry as follows:
\begin{equation}
 \begin{split}
    &[P'_{I}\,,P'_{J}]=0\,,\quad [E'^{I}{}_{J}\,,P'_{K}]=\delta^{I}{}_{K}P'_{J}\,,\quad [E'^{I}{}_{J}\,,E'^{K}{}_{L}]=\delta^{I}{}_{L}E'^{K}{}_{J}-\delta^{K}{}_{J}E'^{I}{}_{L}\,,\\
    &[D'\,,D']=0\,,\quad [D'\,,P'_{I}]=0\,,\quad [D'\,,E'^{I}{}_{J}]=0\,,\\
    &[S'^{I}{}_{J}\,,P'_{K}]=0\,,\quad [S'^{I}{}_{J}\,,E'^{K}{}_{L}]=0\,,\quad [S'^{I}{}_{J}\,,D']=0\,,\quad [S'^{I}{}_{J}\,,S'^{K}{}_{L}]=0\,.
\end{split}
\label{Lie algebra of Riemann-Cartan}
\end{equation}
The algebras in the first line above are nothing but the affine algebra. In the same manner as the cases of the dilation and the shear, the full contraction, {\it i\,.e\,.,} $\epsilon_{\rm \ 0}=(+0\,,+0\,,+0)$ does not lead directly to the vanishing curvature and/or torsion as follows:
\begin{equation}
    T=d\mathbf{e}^{-1}\,,\quad \Omega^{(E)}=d\omega^{(E)}\,.
\label{Riemann and Cartan geometry in abelian affine algebra}
\end{equation}
That is, in order to vanish the torsion and/or curvature 2-forms, some additional conditions are necessary. Using Eq.~(\ref{Relation between Coframe and Translation}), the condition of vanishing torsion is given as follows
\begin{equation}
    d\mathbf{e}^{-1}+d^{2}\zeta=d\mathbf{e}^{-1}:=0\,,
\label{Condition for vanishing torsion in Riemann and Cartan geometry in abelian affine algebra}
\end{equation}
This is nothing but just imposing $T:=0$. That is, the theory is consistent. In the case of the curvature, the so-called Weitzenb\"{o}ch gauge condition
\begin{equation}
   \omega^{(E)}:=0 \,,\quad{\rm or\ equivalently\,,}\quad {\omega^{(E)}}^{I}{}_{\mu J}:=0
\label{Weitzenboch gauge}
\end{equation}
realizes the vanishing curvature. Under the imposition of the above two conditions, the geometry becomes either $E_{n}$ or $M_{n}$. In Sec.~\ref{03}, in particular, the Weitzenb\"{o}ch connection is derived as in Eq.~(\ref{Weitzenboch connection}). Under the imposition of Eq.~(\ref{Weitzenboch gauge}), the connection is given as follows:
\begin{equation}
    \tilde{\Gamma}^{\rho}_{\mu\nu}=e_{I}{}^{\rho}\partial_{\mu}e^{I}{}_{\nu}\,.
\label{Weitzenboch connection in Weitzenboch gauge}
\end{equation}
Using the formula of torsion 2-form in a coordinate/holonomic basis (See Eq.~(\ref{Ricci-like expression of torsion tensor}) in detail), the vanishing torsion is equivalent to the existence of the so-called St\"{u}kelberg fields, denote $\eta=\eta^{I}\xi_{I}$, as follows:
\begin{equation}
    e^{I}{}_{\mu}=\partial_{\mu}\eta^{I}\,,\quad {\rm or\ equivalently,}\quad \mathbf{e}^{-1}=e^{I}\otimes\xi_{I}=d\eta^{I}\otimes\xi_{I}\,.
\label{Veilbein in Stuekelberg fields}
\end{equation}
This, of course, satisfies Eq.~(\ref{Condition for vanishing torsion in Riemann and Cartan geometry in abelian affine algebra}). Therefore, even if the flat geometries $E_{n}$ or $M_{n}$, the internal space still has the degrees of freedom of the St\"{u}kelberg fields. Furthermore, the imposition of the so-called coincident gauge condition~\cite{BeltranJimenez:2017tkd,BeltranJimenez:2022azb,Tomonari:2023wcs,Hu:2023gui}, $\eta^{I}=A^{I}{}_{\mu}x^{\mu}+B^{I}$, where $x^{\mu}$ are the coordinates of a point in the base space $\mathcal{M}$, $A^{I}{}_{\mu}\in G(n\,;\,\mathbb{R})$ (it is a global symmetry), and $B=\xi_{I}B^{I}$ is a constant vector with respect to the global symmetry, then the Weitzenb\"{o}ch connection given in Eq.~(\ref{Weitzenboch connection in Weitzenboch gauge}) vanishes. 

Finally, let us consider the contraction generated by $\epsilon_{\rm \ abelian\ MA}=(+0\,,1\,,1)$. The algebra is given as follows:
\begin{equation}
\begin{split}
    &[P'_{I}\,,P'_{J}]=0\,,\quad [E'^{I}{}_{J}\,,P'_{K}]=0\,,\quad [E'^{I}{}_{J}\,,E'^{K}{}_{L}]=0\,,\\
    &[D'\,,D']=0\,,\quad [D'\,,P'_{I}]=P'_{I}\,,\quad [D'\,,E'^{I}{}_{J}]=0\,,\\
    &[S'^{I}{}_{J}\,,P'_{K}]=\delta^{I}{}_{K}P'_{J}\,,\quad [S'^{I}{}_{J}\,,E'^{K}{}_{L}]=0\,,\quad [S'^{I}{}_{J}\,,D']=0\,,\quad [S'^{I}{}_{J}\,,S'^{K}{}_{L}]=0\,,
\end{split}
\label{Parametrized Lie algebra of abelian metric-affine group}
\end{equation}
where ``{abelian}'' in the subscript of the parameter space means that the algebras in the first line above, which was the Poincare algebra before performing the contraction, turn now to be a set of commutative algebras. Hereinafter, we use this notation. Imposing Eq.~(\ref{Condition for vanishing torsion in Riemann and Cartan geometry in abelian affine algebra}) and Eq.~(\ref{Weitzenboch gauge}), the curvature and the torsion vanish; the geometry turns into the symmetric teleparallel geometry, $S_{n}$, which is described only by the dilation and shear, or equivalently, the non-metricity. Notice that Eq.~(\ref{Weitzenboch connection in Weitzenboch gauge}) and Eq.~(\ref{Veilbein in Stuekelberg fields}) hold for the contraction generated by $\epsilon_{\rm \ abelian\ MA}=(+0\,,1\,,1)$. The dilation-free case, $\epsilon_{\rm \ abelian\ MA\ with\ dilation-free}=(+0\,,+0\,,1)$, and the shear-free case, $\epsilon_{\rm \ abelian\ MA\ with\ shear-free}=(0+\,,1\,,+0)$, can be constituted in the same manner.

\subsection{\label{07:03}An extension of metric-affine algebra and its contraction}
An extension of the Lie algebra of $MA(n\,;\,\mathbb{R})$, let us denote $EMA(n\,;\,\mathbb{R})$, can be given follows:
\begin{equation}
\begin{split}
    &[P_{I}\,,P_{J}]=0\,,\quad [E^{I}{}_{J}\,,P_{K}]={C^{(A3)}}^{IL}{}_{JK}P_{L}\,,\quad [E^{I}{}_{J}\,,E^{K}{}_{L}]={C^{(A6)}}^{IKN}{}_{JLM}E^{M}{}_{N}\,,\\
    &[D\,,D]=0\,,\quad [D\,,P_{I}]={C^{(W4)}}^{J}{}_{I}P_{J}\,,\quad [D\,,E^{I}{}_{J}]={C^{(W9)}}^{I}{}_{J}D\,,\\
    &[S^{I}{}_{J}\,,P_{K}]={C^{(MA1)}}^{IL}{}_{JK}P_{L}\,,\quad [S^{I}{}_{J}\,,E^{K}{}_{L}]={C^{(MA16)}}^{IKM}{}_{JLN}S^{N}{}_{M}\,,\\
    &[S^{I}{}_{J}\,,D]=0\,,\quad [S^{I}{}_{J}\,,S^{K}{}_{L}]=0\,.
\end{split}
\label{Lie algebra of EMA}
\end{equation}
If ${C^{(A3)}}^{IL}{}_{JK}=\delta^{I}{}_{K}\delta^{L}{}_{J}$, ${C^{(A6)}}^{IKN}{}_{JLM}=\delta^{I}{}_{L}\delta^{K}{}_{M}\delta^{N}{}_{J}-\delta^{I}{}_{M}\delta^{K}{}_{J}\delta^{N}{}_{L}$, ${C^{(W4)}}^{J}{}_{I}=\delta^{J}{}_{I}$, ${C^{(MA1)}}^{IL}{}_{JK}=\delta^{I}{}_{K}\delta^{L}{}_{J}$, and otherwise vanish then the Lie algebra $\mathfrak{ema}(n\,;\,\mathbb{R})$ of $EMA(n\,;\,\mathbb{R})$ turns into that of $MA(n\,;\,\mathbb{R})$. Therefore, the algebra of Eq.~(\ref{Lie algebra of EMA}) is an extension of the algebra of Eq.~(\ref{Lie algebra of metric-affine group}). Another important algebra, which includes the Poincare algebra as a subalgebra, is obtained under the following structure constants: ${C^{(A3)}}^{IL}{}_{JK}=\delta^{I}{}_{K}\delta^{L}{}_{J}-g^{IL}{}_{}g^{}{}_{JK}$, ${C^{(A6)}}^{IKN}{}_{JLM}=g^{IK}{}_{}g^{}{}_{JM}\delta^{N}{}_{L}-g^{KN}{}_{}g^{}{}_{JM}\delta^{I}{}_{L}-\delta^{K}{}_{J}\delta^{I}{}_{M}\delta^{N}{}_{L}+g^{KN}{}_{}g^{}{}_{JL}\delta^{I}{}_{M}$, ${C^{(W4)}}^{J}{}_{I}=\delta^{J}{}_{I}$, ${C^{(MA1)}}^{IL}{}_{JK}=\delta^{I}{}_{K}\delta^{L}{}_{J}$, and otherwise vanish. Then $\mathfrak{ema}(n\,;\,\mathbb{R})$ is an extension of the Poincare algebra with the dilation and the shear. Hereinafter, however, we treat the algebra~(\ref{Lie algebra of EMA}) without specifying the structure constants: after obtaining a contraction, we specify a set of structure constants and then develop a theory of gravity, although constructing physical theories is out of scope of the current paper. 

The parametrization given in Eq.~(\ref{Parametrization for IW contraction}) is also taken for the above algebra~(\ref{Lie algebra of EMA}) to perform the contraction. Then the parametrized algebra of Eq.~(\ref{Lie algebra of EMA}) is given as follows:
\begin{equation}
\begin{split}
    &[P'_{I}\,,P'_{J}]=0\,,\quad [E'^{I}{}_{J}\,,P'_{K}]=\epsilon^{(E)}{C^{(A3)}}^{IL}{}_{JK}P'_{L}\,,\quad [E'^{I}{}_{J}\,,E'^{K}{}_{L}]=\epsilon^{(E)}{C^{(A6)}}^{IKN}{}_{JLM}E'^{M}{}_{N}\,,\\
    &[D'\,,D']=0\,,\quad [D'\,,P'_{I}]=\epsilon^{(D)}{C^{(W4)}}^{J}{}_{I}P'_{J}\,,\quad [D'\,,E'^{I}{}_{J}]=\frac{\epsilon^{(D)}\epsilon^{(E)}}{\epsilon^{(P)}}{C^{(W9)}}^{IK}{}_{J}P'_{K}\,,\\
    &[S'^{I}{}_{J}\,,P'_{K}]=\epsilon^{(S)}{C^{(MA1)}}^{IL}{}_{JK}P'_{L}\,,\quad [S'^{I}{}_{J}\,,E'^{K}{}_{L}]=\frac{\epsilon^{(S)}\epsilon^{(E)}}{\epsilon^{(P)}}{C^{(MA16)}}^{IKM}{}_{JL}P'_{M}\,,\\
    &[S'^{I}{}_{J}\,,D']=0\,,\quad [S'^{I}{}_{J}\,,S'^{K}{}_{L}]=0\,.
\end{split}
\label{Parametrized Lie algebra of EMA}
\end{equation}
Notice that this algebra explicitly contains the parameter $\epsilon^{(P)}$. Therefore, we use the parameter space $\epsilon_{\rm \ type\ of\ contraction/geometry\ in\ EMA}=(\epsilon^{(P)}\,,\epsilon^{(E)}\,,\epsilon^{(D)}\,,\epsilon^{(S)})$. This is a different situation from the contraction of the metric-affine algebra. The geometric quantities of the contraction generated by $\epsilon_{\rm \ 0}=(1\,,+0\,,+0\,,+0)$, $\epsilon_{\rm \ abelian\ EMA}=(1\,,+0\,,1\,,1)$, $\epsilon_{\rm \ abelian\ EMA\ with\ dilation-free}=(1\,,+0\,,+0\,,1)$, and $\epsilon_{\rm \ abelian\ EMA\ with\ shear-free}=(1\,,+0\,,1\,,+0)$ are constructed in the same manner as the metric-affine case up to the difference of structure constants (local symmetry). That is, the departures from the metric-affine case are caused by the non-vanishing parameter $\epsilon^{(E)}$. 

Let us consider the geometric quantities in the contraction generated by $\epsilon_{\rm \ EMA}=(1\,,1\,,1\,,1)$. Then the curvature and torsion 2-forms are given as Eq.~(\ref{Riemann and Cartan geometry}). The dilation and shear 2-forms, which are given as Eq.~(\ref{Dilation}) and Eq.~(\ref{Shear}), respectively, are changed as follows:
\begin{equation}
    \Delta=\frac{1}{n}\,{\rm Tr}\,(\,Q\,)\,\mathbf{e}^{-1}+\omega^{(E)}\wedge\,\omega^{(D)}\,
\label{Dilation in EMA}
\end{equation}
and
\begin{equation}
    {\nearrow\!\!\!\!\!\!\!\Delta}=\left(Q-\frac{1}{n}\,{\rm Tr}\,(\,Q\,)\right)\,\mathbf{e}^{-1}+\omega^{(E)}\wedge\,\omega^{(S)}\,,
\label{Shear in EMA}
\end{equation}
respectively. These changes are caused by the violation of the commutativity of $D$ and $E^{I}{}_{J}$, and, $S^{I}{}_{J}$ and $E^{I}{}_{J}$. Of course, the contraction generated by $\epsilon_{\rm \ EMA\ with\ dilation-free}=(1\,,1\,,+0\,,1)$ and $\epsilon_{\rm \ EMA\ with\ shear-free}=(1\,,1\,,1\,,+0)$ leads to $\Delta=0$ and ${\nearrow\!\!\!\!\!\!\!\Delta}=0$ under the imposition of the gauge conditions given in Eq.~(\ref{Gauge-fixing for vanishing dilation}) and Eq.~(\ref{Gauge-fixing for vanishing shear}), respectively. 

The parameter space $\epsilon_{\rm \ type\ of\ contraction/geometry\ in\ EMA}=(0+\,,\epsilon^{(E)}\,,\epsilon^{(D)}\,,\epsilon^{(S)})$ provides a set of new sort of geometric quantities. The coefficients of the third algebra of the second line and that of the second algebra of the third line in the algebra~(\ref{Parametrized Lie algebra of EMA}) diverge unless an appropriate set of the parameters $\epsilon^{(E)}$, $\epsilon^{(D)}$, and $\epsilon^{(S)}$ converges to zero simultaneously. The case of $\epsilon_{\rm \ abelian\ EMA}=(+0\,,+0\,,\epsilon^{(D)}\,,\epsilon^{(S)})$ is the same situation as that of $\epsilon_{\rm \ abelian\ MA}=(+0\,,1\,,1)$ in the metric-affine case up to the difference of structure constants (local symmetry). Therefore, the intriguing cases are those of parameter spaces given by $\epsilon_{\rm \ type\ of\ contraction/geometry\ in\ EMA}=(+0\,,1\,,\epsilon^{(D)}\,,\epsilon^{(S)})$. If the contraction is performed by the parameters $\epsilon_{\rm \ EMA\ with\ shear-free}=(+0\,,1\,,1\,,+0)$ under the imposition of the gauge conditions~(\ref{Gauge-fixing for vanishing shear}) and~(\ref{Gauge-fixing for vanishing dilation}) then the dilation given in Eq.~(\ref{Dilation in EMA}) remains but the shear given in Eq.~(\ref{Shear in EMA}) turns into ${\nearrow\!\!\!\!\!\!\!\Delta}=\omega^{(E)}\wedge\,\omega^{(S)}$. In the case of $\epsilon_{\rm \ EMA\ with\ dilation-free}=(+0\,,1\,,+0\,,1)$ under the same gauge conditions~(\ref{Gauge-fixing for vanishing shear}) and~(\ref{Gauge-fixing for vanishing dilation}), the shear given in Eq.~(\ref{Shear in EMA}) remains but the dilation given in Eq.~(\ref{Dilation in EMA}) turns into $\Delta=\omega^{(E)}\wedge\,\omega^{(D)}$. Remark, here, that the limitations of $\epsilon^{(D)}\rightarrow+0$ and $\epsilon^{(S)}\rightarrow+0$ are taken in the same order magnitude as $\epsilon^{(P)}\rightarrow+0$. The important point here is that these two algebras provide new geometric quantities, which are different from the dilation, the shear, or the non-metricity while holding the richness of the curvature and torsion 2-forms. This richness together with new geometric quantities would go beyond the metric-affine geometry and yield a new class of gauge theories of gravity. 

\subsection{\label{07:04}An extension of de Sitter/anti-de Sitter algebra and its contraction}
Finally, let us consider the following extended dS/AdS Lie algebra:
\begin{equation}
\begin{split}
    &[P_{I}\,,P_{J}]={C^{(A2)}}^{K}{}_{IJL}E^{L}{}_{K}\,,\quad [E^{I}{}_{J}\,,P_{K}]={C^{(A3)}}^{IL}{}_{JK}P_{L}\,,\\
    &[E^{I}{}_{J}\,,E^{K}{}_{L}]={C^{(A6)}}^{IKN}{}_{JLM}E^{M}{}_{N}\,,\\
    &[D\,,D]=0\,,\quad [D\,,P_{I}]=P_{I}\,,\quad [D\,,E^{I}{}_{J}]=0\,,\\
    &[S^{I}{}_{J}\,,P_{K}]=\delta^{I}{}_{K}P_{J}\,,\quad [S^{I}{}_{J}\,,E^{K}{}_{L}]=0\,,\quad [S^{I}{}_{J}\,,D]=0\,,\quad [S^{I}{}_{J}\,,S^{K}{}_{L}]=0\,,
\end{split}
\label{Lie algebra of dS or AdS with dilation and shear}
\end{equation}
where ${C^{(A2)}}^{K}{}_{IJL}=\epsilon\delta^{K}{}_{J}g^{}{}_{IL}$, ${C^{(A3)}}^{IL}{}_{JK}=\delta^{I}{}_{K}\delta^{L}{}_{J}-g^{IL}{}_{}g^{}{}_{JK}$, and ${C^{(A6)}}^{IKN}{}_{JLM}=g^{IK}{}_{}g^{}{}_{JM}\delta^{N}{}_{L}-g^{KN}{}_{}g^{}{}_{JM}\delta^{I}{}_{L}-\delta^{K}{}_{J}\delta^{I}{}_{M}\delta^{N}{}_{L}+g^{KN}{}_{}g^{}{}_{JL}\delta^{I}{}_{M}$. If $\epsilon=+1\,,-1$ then the algebras of the first line become that of the de Sitter space and anti-de Sitter space, respectively~\cite{deSitter:1917,Wise:2006sm,Blagojevic:2012bc,Blagojevic:2013xpa}. The case of $\epsilon=0$ is contained as a special case of the extended metric-affine algebra in the previous Sec.~\ref{07:03}. In order to perform the contraction, the generators $P_{I}$, $E^{I}{}_{J}$, $D$, and $S^{I}{}_{J}$ are parametrized by Eq.~(\ref{Parametrization for IW contraction}). Then the following parametrized Lie algebras are derived:
\begin{equation}
\begin{split}
    &[P'_{I}\,,P'_{J}]=\frac{(\epsilon^{(P)})^{2}}{\epsilon^{(E)}}{C^{(A2)}}^{K}{}_{IJL}E'^{L}{}_{K}\,,\quad [E'^{I}{}_{J}\,,P'_{K}]=\epsilon^{(E)}{C^{(A3)}}^{IL}{}_{JK}P'_{L}\,,\\
    &[E'^{I}{}_{J}\,,E'^{K}{}_{L}]=\epsilon^{(E)}{C^{(A6)}}^{IKN}{}_{JLM}E'^{M}{}_{N}\,,\\
    &[D'\,,D']=0\,,\quad [D'\,,P'_{I}]=\epsilon^{(D)}P'_{I}\,,\quad [D'\,,E'^{I}{}_{J}]=0\,,\\
    &[S'^{I}{}_{J}\,,P'_{K}]=\epsilon^{(S)}\delta^{I}{}_{K}P'_{J}\,,\quad [S'^{I}{}_{J}\,,E'^{K}{}_{L}]=0\,,\quad [S'^{I}{}_{J}\,,D]=0\,,\quad [S'^{I}{}_{J}\,,S'^{K}{}_{L}]=0\,.
\end{split}
\label{Parametrized Lie algebra of dS or AdS with dilation and shear}
\end{equation}
The contraction of the above algebra generated by $\epsilon_{\rm \ Poincare\ in\ extended\ dS/AdS}=(+0\,,1\,,\epsilon^{(D)}\,,\epsilon^{(S)})$ results in that of the algebra $\mathfrak{ema}(n\,,\mathbb{R})$ generated by $\epsilon_{\rm \ Poincare\ in\ EMA}=(\epsilon^{(P)}\,,1\,,\epsilon^{(D)}\,,\epsilon^{(S)})$, where $\epsilon^{(P)}$, $\epsilon^{(D)}$, and $\epsilon^{(S)}$ can be taken as an arbitrary value in $(0\,,1]$. So is for the contraction of algebras between that generated by $\epsilon_{\rm \ abelian\ Poincare\ in\ extended\ dS/AdS}=(+0\,,+0\,,\epsilon^{(D)}\,,\epsilon^{(S)})$ in the extended dS/AdS algebra and that generated by $\epsilon_{\rm \ abelian\ Poincare\ in\ EMA}=$\\
$(\epsilon^{(P)}\,,+0\,,\epsilon^{(D)}\,,\epsilon^{(S)})$ in the extended metric-affine algebra under the choice of the structure constants as the Poincare algebra. Remark, here, that the contractions generated by $\epsilon_{\rm \ type\ of\ contraction/geometry\ in\ extended\ dS/AdS}=(1\,,+0\,,\epsilon^{(D)}\,,\epsilon^{(S)})$ do not exist due to the divergence of the factor $(\epsilon^{(P)})^{2}/\epsilon^{(E)}$ of the first algebra in the first line. Therefore, it is enough to consider the case of $\epsilon_{\rm \ type\ of\ contraction/geometry\ in\ extended\ dS/AdS}=(1\,,1\,,\epsilon^{(D)}\,,\epsilon^{(S)})$.

The contraction generated by $\epsilon_{\rm \ type\ of\ contraction/geometry\ in\ extended\ dS/AdS}=(1\,,1\,,\epsilon^{(D)}\,,\epsilon^{(S)})$ alternates the curvature 2-form given in Eq.~(\ref{Riemann and Cartan geometry}) as follows
\begin{equation}
    \Omega^{(E)}=d_{\nabla}\omega^{(E)}=d\omega^{(E)}+\omega^{(E)}\wedge \omega^{(E)}+\mathbf{e}^{-1}\wedge \mathbf{e}^{-1}\,.
\label{dS/AdS curvature 2-forms}
\end{equation}
The torsion 2-form given in Eq.~(\ref{Riemann and Cartan geometry}) does not change. So are the dilation and shear 2-forms, which are given in Eq.~(\ref{Dilation}) and Eq.~(\ref{Shear}), respectively. Notice that a new geometrical quantity arises; the above quantity is not the curvature 2-form in the sense of the metric-affine geometry due to the existence of the third term. That is, this departure in Eq.~(\ref{dS/AdS curvature 2-forms}) from the metric-affine geometry would yields a new class of gauge theories of gravity. In addition, this geometry is not just the (anti-)de Sitter geometry due to the existence of the dilation and the shear (, or equivalently, the non-metricity).

\subsection{\label{07:05}A physical application of In\"{o}n\"{u}-Wigner contraction to cosmological constant}
Finally, let us illustrate briefly a relation to general relativity as an example in physics. It is known that the third term in Eq.~(\ref{dS/AdS curvature 2-forms}) relates to the so-called cosmological constant in general relativity if the torsion, the dilation, and the shear vanish~\cite{Wise:2006sm,MacDowell:1977jt}. In order to realize this, of course, it is enough to perform further contraction generated by $\epsilon_{\rm \ dS/AdS}=(1\,,1\,,+0\,,+0)$ and to impose the torsion-free condition given in Eq.~(\ref{Condition for vanishing torsion in Riemann and Cartan geometry in abelian affine algebra}) and the gauge conditions~(\ref{Gauge-fixing for vanishing shear}) and~(\ref{Gauge-fixing for vanishing dilation}). In addition, the contraction generated by $\epsilon_{\rm \ V_{n}}=(+0\,,1\,,+0\,,+0)$ removes out the third term in Eq.~(\ref{dS/AdS curvature 2-forms}), and the geometry results in the (pseudo-)Riemann geometry; this is nothing but the contraction of the metric-affine geometry generated by $\epsilon_{\rm \ V_{n}}=(1\,,+0\,,+0)$. In this way, utilizing the methodology established throughout the current paper, geometric quantities, gauge structures, and geometries are related together in the systematic manner, and it would give insight into clarifying the physically intriguing models in the metric-affine gravity.

\section{\label{08}Conclusions}
In this paper, based on the principal bundle theory, a unified-description of the curvature, torsion, and non-metricity 2-forms was investigated by extending the M\"{o}bius representation and Cartan connection of the Riemann-Cartan geometry. After that, the In\"{o}n\"{u}-Wigner contraction was introduced and the correspondences between each geometry and algebra of each gauge group were clarified. Then the dilation and shear 2-forms, or equivalently, the non-metricity 2-form, dropped out by performing the contraction under the imposition of appropriate gauge conditions. So did the curvature and torsion 2-forms. The contraction led also to the possibilities for introducing new geometric quantities by extending the metric-affine algebra in the way that non-commutative algebras are appended by manipulating the structure constants.  Finally, extending the (anti-)de Sitter algebra, the dilation and shear 2-forms were unified as the geometric quantities to describe the corresponding geometry: the (anti-)de Sitter space with the dilation and torsion 2-forms (, or equivalently, non-metricity 2-form). 

As mentioned in Sec.~\ref{01}, the condition of vanishing non-metricity demands a gauge condition in the internal space, and, if it is the case, unveiling such conditions should be investigated. One such case was to choose the trivial gauge. Other possibilities to realize this situation will appear in a sequel paper. Furthermore, in the cases of geometries obtained by alternating the gauge group from the metric-affine group to other groups, the emergence of new geometric quantities would imply that curvature, torsion, and non-metriciy 2-forms would need some additional gauge conditions for vanishing. Or, some of these quantities might describe physical phenomena. The (anti-)de Sitter geometry was such a case. To unveil these conditions in detail or the possibilities for describing physical phenomena would play a crucial role in constructing physically valid theories of gravity. After completing these investigations, the action functionals in terms of these quantities should be composed, and the Dirac-Bergmann analysis should also be performed~\cite{Dirac:1950pj,Dirac:1958sc,Dirac:1958sq,Anderson:1951,Bergmann:1949a,Bergmann:1949b,Bergmann:1950}. In particular, the latter analysis is important to unveil the number of possible propagating degrees of freedom, the existence of (Ostrogradski's) ghost degrees of freedom~\cite{Ostrogradsky:1850fid,Woodard:2015zca}, and the Dirac structure: the existence of constraints and those classifications into first- and second-class, and that of gauge symmetries. These quests are left for future works. 

As shown throughout the current paper, the fertility of the gauge theories of gravity should be ascribed to the structure of the internal space. The authors in Ref~\cite{Gomes:2023hyk} unveiled that teleparallel gravity for the sake of cosmology, meaning that the geometry is homogeneous and isotropic, has five possible branches in the theory. This sort of classification play an important role in relating the number of propagating degrees of freedom revealed by the Dirac-Bergmann analysis to that number speculated by the cosmological perturbation. In fact, the authors in Ref~\cite{Blagojevic:2020dyq} unveiled that the analysis of $f(T)$-gravity indicates the existence of five possible sectors and each sector generically has different propagating degrees of freedom~\footnote{
In particular, Ref.~\cite{Ferraro:2018tpu} unveiled that $f(T)$-gravity has only one extra degrees of freedom in the comparision to the case of general relativity and, at a glance, it is inconsistent with the common knowledge that $f(T)$-gravity has five degrees of freedom in four-dimensional spacetime. In this point, Ref.~\cite{Tomonari:2023wcs} unveiled that the analysis in Ref.~\cite{Ferraro:2018tpu} is nothing but that in a generic sector which is also investigated in Ref.~\cite{Blagojevic:2020dyq}. In detail, see Sec.V-A in Ref.~\cite{Tomonari:2023wcs}. 
}.
The relationships between the branches and the sectors are not unveiled today. Similarly, on one hand, the authors in Ref~\cite{Hu:2022anq} unveiled that the analysis of coincident $f(Q)$-gravity indicates the existence of eight propagating degrees of freedom in a sector of the theory. On the other hand, the authors in Ref~\cite{Tomonari:2023wcs} showed that the theory has six propagating degrees of freedom in another sector of the theory, together with a possibility that the number is possibly seven or five in other sectors. From the viewpoint of cosmological perturbation, the authors in Ref~\cite{Gomes:2023tur} unveiled that seven propagating degrees of freedom exist in the non-trivial branch I. For the trivial and non-trivial branch II, the number could be four or less. The relationships between the branches and the sectors are also not unveiled today. In this regard, just scrutinizing curvature, torsion, and non-metricity themselves on the tangent bundle, which is likely to be a standard formalism in the present days, may be oversimplified. The fertility should be scrutinized in detail. 

Finally, we briefly mention the possible extension of the method that is presented in the current paper from the viewpoints of $f(R)$-, $f(T)$-, and $f(Q)$-gravity. There are comprehensive reviews on each extended/modified theory of gravity: Refs.~\cite{DeFelice:2010aj,Nojiri:2006ri}, Ref.~\cite{Cai:2015emx}, and Refs.~\cite{Heisenberg:2023lru,Zhao:2021zab}, respectively. A common review on the generic framework of these theories is given in Ref.~\cite{Capozziello:2011et}. As mentioned in the previous paragraph, generically, the extended/modified theories of gravity have several sectors, and each sector has a proper constraint structure. This would lead to the proper physical phenomena for each sector. The crucial point for this bifurcation is symmetry breaking. GR has two fundamental symmetries: the local Lorentz symmetry, or generically speaking, the frame invariance, and the diffeomorphism symmetry. These symmetries are represented in terms of the Dirac analysis by the corresponding PB-algebra of the structure group $G$ of the internal bundle (See Sec.~\ref{03:01} and~\ref{02}) and the Hypersurface Deformation Algebra (HDA)~\cite{Dirac:1958sc} of the tangent bundle of a given spacetime manifold. $f(R)$-gravity holds these two sets of PB-algebras. This implies that the theory only has a single sector. In the case of $f(T)$-gravity, however, this theory loses the local Lorentz invariance, and this causes the bifurcation, although the HDA holds. Therefore, investigating further the pattern of the symmetry breaking, we would identify the relationships between the branches and the sectors mentioned in the previous paragraph. Then, we can perform the In\"{o}n\"{u}-Wigner contraction that corresponds to the breaking pattern of the PB-algebra and would find a comprehensive construction of the theory. It would be in the same situation as $f(Q)$-gravity. Furthermore, since the method presented in the current paper can extend the structure group freely to another one, it might be extended/modified to a theory beyond MAG. All of these investigations would be great future works. 

\section*{Acknowledgments}
KT would like to thank all colleagues of the cosmology theory group in Tokyo Institute of Technology. KT is supported by Tokyo Tech Fund Hidetoshi Kusama Scholarship.
\section*{Declarations}
\section*{Data availability}
Data sharing not applicable to this article as no datasets were generated or analysed during the current study.
\section*{Conflicts of interests}
The author have no competing interests to declare that are relevant to the content of this article.

\appendix
\section{\label{02}Reconstruction of metric-affine geometry: curvature, torsion, and non-metricity 2-forms}
In this appendix, fundamental geometric quantities, {\it i\,.e\,.,} curvature, torsion, and non-metricity 2-forms, of the metric-affine geometry based on a bundle theory are introduced in a self-contained manner, for the purpose of reformulating these quantities in terms of gauge theory. Let us assume that a $G$-bundle: $\mathfrak{G}=(\mathcal{E}\,,\mathcal{M}\,,\pi\,,\mathcal{F}\,,G)$ exists, where $\mathcal{E}$ is a total space, $\mathcal{M}$ is a base manifold, $\pi$ is a differentiable onto map from $\mathcal{E}$ to $\mathcal{M}$, $\mathcal{F}$ is a standard fiber, and $G$ is a structure/gauge group. 

\subsection{\label{02:01}Principal G-bundle, Ehresmann connection, and curvature 2-form}
A principal $G$-bundle is a $G$-bundle $\mathfrak{G}$ equipped with a right action of a group $G$, {\it i\,.e\,.,} $R:\,\mathcal{P}\times G\rightarrow \mathcal{P};\,(u\,,g)\mapsto u\cdot g$, where ``$\,\cdot\,$'' denotes the right action of the group $G$, acting on the total space $\mathcal{P}$ in the following manner: $(u\cdot g)g'=u\cdot(gg')$ and $u\cdot g=u$, and it satisfies the following conditions; (i) $\pi(u\cdot g)=\pi(u)$; (ii) $\pi(u)=\pi(u')$ implies the existence of an element $g\in G$ such that $u'=u\cdot g$ (simply transitivity). In addition, assume that (iii) All sections $\sigma_{i}:\,U_{i}\rightarrow\pi^{-1}(U_{i})=U_{i}\times G$ are differentiable, where $U_{i}\in\mathfrak{U}$ and $\mathfrak{U}$ is an open covering of the base space $\mathcal{M}$. Conditions (i) and (ii) identify the fiber $\mathcal{F}$ with the structure group $G$. Therefore, the principal $G$-bundle is denoted as $\mathfrak{P}=(\mathcal{P}\,,\mathcal{M}\,,\pi\,,G\,;\,R)$. Condition (iii) leads to a local trivialization $\varphi_{i}:\,\mathcal{P}|_{U_{i}}\rightarrow U_{i}\times G;\,\sigma_{i}(p)\mapsto (p\,,g)$ for $U_{i}\in\mathfrak{U}$. 
The transition functions are defined by using the local trivializations, $\{\varphi_{i}\}_{i\in I}$, with respect to the {\it right} action of the structure group $G$ as usual: $\{\tau_{ij}=\varphi^{-1}_{i}\circ\varphi_{j}\}$. 

Then, the Ehresmann connection is introduced for the principle $G$-bundle $\mathfrak{P}$ as follows~\cite{Ehresmann:1952,Kobayashi:1989,Kobayashi:1996,Sharpe:1997}:
\begin{equation}
\begin{split}
    &{\rm (i)}\quad \omega(A^{*})=A\quad (A\in\mathfrak{g})\,,\\
    &{\rm (ii)}\quad R_{g}{}^{*}\omega(X)=Ad_{g}(\omega)=g^{-1}\omega g\,,
\end{split}
\label{Ehresmann connection}
\end{equation}
where $\mathfrak{g}$ is the Lie algebra of the structure group $G$, $\omega$ is a $\mathfrak{g}$-valued $1$-form on $\mathcal{P}$, $R_{g}{}^{*}$ is the pull back operator of the right action with respect to an element $g\in G$, and $Ad_{g}$ is the adjoint representation with respect to $g\in G$. $A^{*}$ is the so-called fundamental vector field defined as follows:
\begin{equation}
    A^{*}{}_{u}f(u)=\left.\frac{d}{dt}\left[f\left(ug(t)\right)\right]\right|_{t=0}\,,
\label{Fundamental vector field}
\end{equation}
where $u\in P$, $g(t)={\rm exp}(tA)\in G$, and $f$ is an arbitrary function on $\mathcal{P}$. The right action, $R\,:\,\mathcal{P}\times G\rightarrow\mathcal{P}$, induces the differential map $dR:\,T\mathcal{P}\times TG\rightarrow T\mathcal{P}$; it means that the right action of $\mathfrak{g}=T_{e}G$ to $\mathcal{P}$ is well-defined. Therefore, the fundamental vector field can be written in the specific form as follows:
\begin{equation}
    A^{*}{}_{u}=uA\,,
\label{Specific form of fundamental vector field}
\end{equation}
and it, of course, belongs to $T_{u}\mathcal{P}$. 

The Ehresmann connection has two different formulations: the geometric and algebraic formulations~\cite{Kobayashi:1989,Kobayashi:1996}. The geometric formulation is given as follows: a $\mathfrak{g}$-valued $1$-form $\omega$ on $\mathcal{P}$ is the Ehresmann connection if and only if the following conditions are satisfied:
\begin{equation}
\begin{split}
    &{\rm (i)}\quad T_{u}\mathcal{P}=V_{u}\mathcal{P}\oplus H_{u}\mathcal{P}\,,\\
    &{\rm (ii)}\quad R_{g}{}^{*}(H_{u}\mathcal{P})=H_{ug}\mathcal{P}\,,
\end{split}
\label{Geometrical Ehresmann connection}
\end{equation}
where $V_{u}\mathcal{P}$ and $H_{u}\mathcal{P}$ are the vertical subspace and the horizontal subspace, respectively, which are defined as follows:
\begin{equation}
    V_{u}\mathcal{P}=\{X\in T_{u}\mathcal{P}\,|\,\pi_{*}(X)=0\}\,,\quad H_{u}\mathcal{P}=\{X\in T_{u}\mathcal{P}\,|\,\omega(X)=0\}\,,
\label{Definition of horizontal and vertical subspace}
\end{equation}
where $\pi_{*}$ is the push forward operator of the projection map $\pi$. In particular, the vertical subspace is expressed equivalently as follows:
\begin{equation}
    V_{u}\mathcal{P}=\{A^{*}_{u}=uA\in T_{u}\mathcal{P}\,|\,A\in\mathfrak{g}\}\,.
\label{Equivalent expression of vertical subspace}
\end{equation}
The equivalence of the two definitions~(\ref{Ehresmann connection}) and~(\ref{Geometrical Ehresmann connection}) is obvious from the definitions of the horizontal and vertical subspaces. Since the principal $G$-bundle $\mathcal{P}$ is expressed as a direct product $\pi^{-1}(U)=U\times G$ in a local region, the above properties give a geometrical interpretation that $U$ and $G$ represent the horizontal and vertical directions on $\pi^{-1}(U)\subset \mathcal{P}$, respectively. 

The curvature/field strength of the Ehresmann connection in the geometric representation~(\ref{Geometrical Ehresmann connection}) is introduced as follows:
\begin{equation}
    \Omega(X\,,Y)=d\omega(X^{H}\,,Y^{H})\,,
\label{Curvature in geometrical representation}
\end{equation}
where $\Omega$ is a $\mathfrak{g}$-valued 2-form on $\mathcal{P}$, $X$ and $Y$ belong to $T\mathcal{P}$, and $X^{H}$ and $Y^{H}$ belong to $H\mathcal{P}$. Expanding this, the following equation is derived:
\begin{equation}
    \Omega(X\,,Y)=-\frac{1}{2}\omega\left([X^{H}\,,Y^{H}]\right)\,.
\label{}
\end{equation}
Therefore, the flatness, which is of course expressed as $\Omega=0$, of the principal $G$-bundle $\mathcal{P}$ is equivalent to satisfying the following condition:
\begin{equation}
    [X^{H}\,,Y^{H}]=0\,.
\label{}
\end{equation}
This condition means that the horizontal subspace $H\mathcal{P}=\bigsqcup_{u\in\mathcal{P}}H_{u}\mathcal{P}$ is Frobenius integrable, and it restores the entire of the total space $\mathcal{P}$; in other words, if $\Omega\neq0$, the Frobenius theorem restores only a part of the total space $\mathcal{P}$ as a submanifold. Notice that the vertical subspace $V\mathcal{P}=\bigsqcup_{u\in\mathcal{P}}V_{u}\mathcal{P}$ is always Frobenius integrable, and it restores the total space $\mathcal{P}$. 

Finally, let us introduce the algebraic formulation of the Ehresmann connection. This formulation plays a crucial role in establishing gauge theories. The algebraic formulation is given as follows: a section $\sigma_{i}:\,U_{i}\rightarrow\pi^{-1}(U_{i})$, where $U_{i}\in\mathfrak{U}$, provides the pull back of the Ehresmann connection $\omega$ in Eq~(\ref{Ehresmann connection}) as follows:
\begin{equation}
    \omega_{i}=\sigma^{*}_{i}\omega\,.
\label{Ehresmann connection in local}
\end{equation}
Then the following relation is satisfied:
\begin{equation}
    \omega_{j}=\tau_{ij}{}^{-1}\omega_{i}\tau_{ij}+\tau_{ij}{}^{-1}d\tau_{ij}\,,
\label{Algebraic Ehresmann connection}
\end{equation}
where $\tau_{ij}\in G$ is the transition function on $U_{i}\cup U_{j}$ for $U_{i}\,,U_{j}\in\mathfrak{U}$. Conversely, if the relation~(\ref{Algebraic Ehresmann connection}) are satisfied then the Ehressmann connection in Eq~(\ref{Ehresmann connection}) is concluded.

The curvature/field strength of the Ehresmann connection in the algebraic formulation~(\ref{Algebraic Ehresmann connection}) is introduced as follows:
\begin{equation}
    \Omega_{i}=d\omega_{i}+\frac{1}{2}[\omega_{i}\,,\omega_{i}]\,.
\label{Curvature in local representation}
\end{equation}
This expression can be derived by calculating the pull back of the definition~(\ref{Curvature in geometrical representation}) by a section $\sigma_{i}$ and is now defined on $\mathfrak{g}$-valued 2-form on $\mathcal{M}$. For the curvature $\Omega_{j}$ on $U_{j}$, the following relation is easily verified:
\begin{equation}
    \Omega_{j}=Ad_{\tau_{ij}}(\Omega_{i})=\tau_{ij}{}^{-1}\Omega_{i}\tau_{ij}\,.
\label{Gauge transformation of curvature 2-form}
\end{equation}
These quantities also live as $\mathfrak{g}$-valued 2-forms on $\mathcal{M}$. 

\subsection{\label{02:02}Curvature and torsion 2-forms}
Let us introduce the curvature and torsion 2-forms of a manifold $\mathcal{M}$ by using a principle $G$-bundle of which base manifold is diffeomorphic to the manifold $\mathcal{M}$. Each fiber $T_{p}\mathcal{M}$ of the tangent bundle $T\mathcal{M}$ has a basis $\{e_{I}\}$, where $p\in\mathcal{M}$ and $I\in\{1\,,2\,,\cdots\,,n\}$. $n$ is the dimension of each fiber, or equivalently, that of the manifold $\mathcal{M}$. Then an isomorphism from $\mathbb{R}^{n}$ to $T_{p}\mathcal{M}$, {\it i\,.e\,.,} $u\,:\,\mathbb{R}^{n}\rightarrow \left<e_{1}\,,e_{2}\,,\cdots\,,e_{n}\right>=T_{p}\mathcal{M}$ always exists, where $\left<\,\cdots\,\right>$ denotes a vector space spanned by the basis ``$\,\cdots\,$''. Gathering all such $u$, denotes the set as ${L(\mathcal{M})}_{p}$, and then construct a disjoint union as follows: ${L(\mathcal{M})}=\bigsqcup_{p\in\mathcal{M}}{L(\mathcal{M})}_{p}$. For each ${L(\mathcal{M})}_{p}$, a right action $R$ of $GL(n\,;\,\mathbb{R})$ is defined in a well-defined manner as follows: $R_{g}\,:\,{L(\mathcal{M})}_{p}\times GL(n\,;\,\mathbb{R})\rightarrow{L(\mathcal{M})}_{p}\,;\,(u\,,g)\mapsto u\cdot g$ since $R_{g}(u)=u\cdot g$ is nothing but a basis transformation of $T_{p}\mathcal{M}$. Then a manifold structure is inserted into the disjoint union ${L(\mathcal{M})}$ by using the inverse map of a local trivialization $\varphi_{U}\,:\,{L(\mathcal{M})}|_{U}\rightarrow U\times GL(n\,;\,\mathbb{R})\,;\,\sigma_{U}(p)\cdot g\rightarrow (p\,,g)$, where $\sigma_{U}$ is a set of a basis of $T_{p}\mathcal{M}$ at point $p\in U$. Based on these structures, the disjoint union ${L(\mathcal{M})}$ becomes a principal $GL(n\,;\,\mathbb{R})$-bundle associated to the tangent bundle $T\mathcal{M}$: $\mathfrak{L}=({L(\mathcal{M})}\,,\mathcal{M}\,,\iota\,,GL(n\,;\,\mathbb{R})\,;\,R)$, where the projection map is given as follows: $\iota\,:\,{L(\mathcal{M})}\rightarrow\mathcal{M}\,;\,{L(\mathcal{M})}|_{p}\mapsto p$. 

From Section~\ref{02:01}, the associated principal bundle ${L(\mathcal{M})}$ has, on one hand, the Ehresmann connection $\omega$, which is a section of $\mathfrak{gl}(n\,;\,\mathbb{R})\otimes T^{*}L(\mathcal{M})$, is defined as Eq.~(\ref{Ehresmann connection}). This 1-form $\omega$ lives in the dual space of the vertical space $VL(\mathcal{M})$. On the other hand, the associated principal bundle ${L(\mathcal{M})}$ has its proper $\mathbb{R}^{n}$-valued 1-forms on ${L(\mathcal{M})}$, denote $\theta^{I}$, defined as follows:
\begin{equation}
    \theta^{I}(X)=\sigma^{-1}_{U}(p)(\iota_{*}X)\,,
\label{Dual basis of L(M)}
\end{equation}
where $X$ is a vector field on ${L(\mathcal{M})}$ and $\sigma_{U}$ is a section of ${L(\mathcal{M})}$. Then the pull back of $\theta^{I}$ becomes the dual basis of a basis of a tangent space $T_{p}\mathcal{M}$: $\sigma_{U}^{*}\theta^{I}(e_{J})=\delta^{I}{}_{J}$, where $\{e_{I}\}$ denotes a basis of $T_{p}\mathcal{M}$. Therefore, these 1-forms $\theta^{I}$ lives in the dual space of the horizontal space $HL(\mathcal{M})$. Remark that the existence of such 1-forms is a proper characteristic of associated principle bundles since generic principle bundles do not have any basis of those fibers. 

For the Ehresmann connection $\omega$, the curvature 2-form is defined by Eq.~(\ref{Curvature in geometrical representation}). Similarly, for the 1-forms $\theta^{I}$, a $\mathbb{R}^{n}$-valued 2-form $\theta^{I}$ is defined as follows:
\begin{equation}
    T(X\,,Y)=d\theta(X^{H}\,,Y^{H})\,,
\label{Torsion in geometrical representation}
\end{equation}
where $X\,,Y$ are sections of $TL(\mathcal{M})$, $\theta\in\left<\theta^{I}\right>$, and $X^{H}\,,Y^{H}\in HL(\mathcal{M})$. This quantity is called the torsion 2-form of the manifold $\mathcal{M}$. In the same manner as the curvature 2-form given in Eq.~(\ref{Curvature in local representation}), the torsion 2-form has the algebraic expression given as follows:
\begin{equation}
    \sigma^{*}_{U}T=d\sigma^{*}_{U}\theta+\frac{1}{2}[\sigma^{*}_{U}\omega\,,\sigma^{*}_{U}\theta]\,.
\label{Torsion in local representation}
\end{equation}
This relation holds on the spacetime manifold $\mathcal{M}$. Where $U$ is an open set of $\mathcal{M}$. Since the curvature 2-form $\Omega$ given in Eq~(\ref{Curvature in geometrical representation}) and the torsion 2-form given in the above equation vanish for $X\,,Y\in VL(\mathcal{M})$, these quantities are expressed in algebraic forms as follows:
\begin{equation}
    \Omega=\Omega^{I}{}_{J}e_{I}\otimes\theta^{J}
\label{Curvature in components expresion in MAG}
\end{equation}
and 
\begin{equation}
    T=\Theta^{I}\otimes e_{I}\,,
\label{Torsion in components expresion in MAG}
\end{equation}
respectively, where $\Omega^{I}{}_{J}$ and $\Theta^{I}$ are defined as follows:
\begin{equation}
    \Omega^{I}{}_{J}=\frac{1}{2}R^{I}{}_{JKL}\theta^{K}\wedge\theta^{L}
\label{Components of curvature in MAG}
\end{equation}
and
\begin{equation}
    \Theta^{I}=\frac{1}{2}T^{I}{}_{JK}\theta^{J}\wedge\theta^{K}\,,
\label{Components of torsion in MAG}
\end{equation}
respectively. Eq.~(\ref{Curvature in components expresion in MAG}) and Eq.~(\ref{Torsion in components expresion in MAG}) are a $\mathfrak{gl}(n\,;\,\mathbb{R})$-valued 2-form on $L(\mathcal{M})$ and a $T\mathcal{M}$-valued 2-form on $L(\mathcal{M})$, respectively. 

Now, let us apply the framework given in Section~\ref{02:01} to the associated principle bundle $L(\mathcal{M})$ and its geometric quantities $\Omega$ and $T$. For the representation $\rho=id$, the principal bundle $L(\mathcal{M})$ turns back into the original vector bundle $T\mathcal{M}$. Then the Ehresmann connection $\omega$ becomes a section of ${\rm End}(T\mathcal{M})$-valued 1-form on $\mathcal{M}$; this connection $\omega$ is nothing but the so-called affine connection~\cite{Cartan:1923,Cartan:1951,Kobayashi:1989,Kobayashi:1996} and usually denoted as `` $\tilde{\Gamma}$ '' when it is necessary to distinguish these connections clearly. Then, for the curvature 2-form, on one hand, it becomes a section of ${\rm End}(T\mathcal{M})$-valued 2-form on $\mathcal{M}$, and the Ricci identity
\begin{equation}
    R(X\,,Y)\xi=\frac{1}{2}\left(\nabla_{X}\nabla_{Y}-\nabla_{Y}\nabla_{X}-\nabla_{[X\,,Y]}\right)\xi\,,
\label{Ricci identity}
\end{equation}
and the structure equation
\begin{equation}
\begin{split}
    &Re_{I}=\Omega_{I}{}^{J}\otimes e_{J}\,,\\
    &\Omega_{I}{}^{J}=d\omega_{I}{}^{J}+\omega_{K}{}^{J}\wedge\omega_{I}{}^{K}=d_{D}\omega_{I}{}^{J}\,,
\end{split}
\label{Structure equation}
\end{equation}
hold, respectively~\cite{Kobayashi:1989,Kobayashi:1996}. On the other hand, for the torsion 2-form, it becomes 
a section of $T\mathcal{M}$-valued 2-form on $\mathcal{M}$, and the resemble formula to the curvature 2-form are derived as follows:
\begin{equation}
    \tilde{T}(X\,,Y)=\frac{1}{2}\left(\tilde{\nabla}_{X}Y-\tilde{\nabla}_{Y}X-[X\,,Y]\right)\,,
\label{Ricci-like expression of torsion tensor}
\end{equation}
where $X$ and $Y$ are vector fields on $\mathcal{M}$, {\it i\,.e\,.,} sections of $T\mathcal{M}$~\cite{Kobayashi:1989,Kobayashi:1996}. Hereinafter, we denote the quantities in the affine connection putting `` $\tilde{\,}$ '' over those notations. Calculating it for a basis $\{e_{I}\}$ of $T\mathcal{M}|_{U}$, where $U$ is an open set of $\mathcal{M}$, the following equations are derived:
\begin{equation}
    \tilde{T}(e_{I}\,,e_{J})=\left(d\theta^{K}+\omega^{K}{}_{L}\wedge\theta^{L}\right)(e_{I}\,,e_{J})\otimes e_{K}\,,
\label{Torsion in component-form 1}
\end{equation}
where $\theta^{I}$ are the dual basis of $e_{I}$; these $\theta^{I}$ correspond to Eq.~(\ref{Dual basis of L(M)}). Comparing the above equation with Eq.~(\ref{Torsion in components expresion in MAG}), the components $\Theta^{I}$ are determined as follows:
\begin{equation}
    \tilde{\Theta}^{I}=d\theta^{I}+\omega_{J}{}^{I}\wedge \theta^{J}=d_{\tilde{\nabla}}\theta^{I}\,,
\label{Torsion in component-form 2}
\end{equation}
where $\omega_{I}{}^{J}$ are the components of the Ehresmann connection $\omega$ and $d_{\nabla}$ is the exterior covariant derivative of the connection $\omega$. This equation is called Cartan's first structure equation~\cite{Cartan:1951,Kobayashi:1989,Kobayashi:1996}. The equation of the curvature 2-form, {\it i\,.e\,.,} 
\begin{equation}
    \tilde{\Omega}_{I}{}^{J}=d\omega_{I}{}^{J}+\omega_{K}{}^{J}\wedge\omega_{I}{}^{K}=d_{\tilde{\nabla}}\omega_{I}{}^{J}\,,
\label{}
\end{equation}
is called Cartan's second structure equation~\cite{Cartan:1951,Kobayashi:1989,Kobayashi:1996}.

In the existence of the torsion, in addition to the Bianchi identity 
\begin{equation}
    d_{\nabla}R=0\,,
\label{Bianchi identity}
\end{equation}
another identity that relates the curvature to the torsion exists:
\begin{equation}
    d_{\tilde{\nabla}}\tilde{T}=\tilde{R}e_{I}\wedge \theta^{I}\,.
\label{Bianchi first identity 1}
\end{equation}
This identity is defined as a section of $T\mathcal{M}\otimes\bigwedge^{3}T^{*}\mathcal{M}$. Acting it to three vector fields $X$, $Y$, and $Z$, the above identity is expressed as follows:
\begin{equation}
    \tilde{R}(X\,,Y)Z+\tilde{R}(Y\,,Z)X+\tilde{R}(Z\,,X)Y=2(d_{\tilde{\nabla}}\tilde{T})(X\,,Y\,,Z)\,.
\label{Bianchi first identity 2}
\end{equation}
These identities~(\ref{Bianchi first identity 1}) and~(\ref{Bianchi first identity 2}) are called Bianchi's first identity~\cite{Kobayashi:1989,Kobayashi:1996}. The Bianchi identity given in Eq.~(\ref{Bianchi identity}) with replacing $d_{\nabla}$ by $d_{\tilde{\nabla}}$ is then called Bianchi's second identity, distinguishing from the first one. 

\subsection{\label{02:03}Non-metricity 2-form}
Let us introduce the non-metricity 2-form~\cite{Kobayashi:1989,Kobayashi:1996}. Assume that a metric tensor $g=g_{IJ}\theta^{I}\otimes\theta^{J}$ of the manifold $\mathcal{M}$ is given. Then the metric tensor generically satisfies the following relation:
\begin{equation}
    dg(X\,,Y)=(\tilde{\nabla} g)(X\,,Y)+g(\tilde{\nabla} X\,,Y)+g(X\,,\tilde{\nabla} Y)\,,
\label{Non-metricity tensor in MAG 1}
\end{equation}
or equivalently, 
\begin{equation}
    Zdg(X\,,Y)=(\tilde{\nabla}_{Z} g)(X\,,Y)+g(\tilde{\nabla}_{Z} X\,,Y)+g(X\,,\tilde{\nabla}_{Z} Y)\,,
\label{Non-metricity tensor in MAG 2}
\end{equation}
where $X$, $Y$, and $Z$ are vector fields on $\mathcal{M}$. Then the quantity $\tilde{\nabla} g$, or equivalently, $\tilde{\nabla}_{X} g$ is defined as the non-metricity 2-form and denoted as $\tilde{Q}$, or equivalently, $\tilde{Q}_{X}$. For the basis $\{e_{I}\}$, and it is expressed as follows:
\begin{equation}
    \tilde{Q}=d_{\tilde{\nabla}}g\,,
\label{Non-metricity tensor from potential}
\end{equation}
or equivalently,
\begin{equation}
\begin{split}
    &\tilde{Q}_{IJ}=\tilde{\nabla}g_{IJ}=dg_{IJ}-\omega_{I}{}^{K}g_{KJ}-\omega_{J}{}^{K}g_{IK}\,,\\
    &\tilde{Q}_{KIJ}=\tilde{\nabla}_{K} g_{IJ}=dg_{IJ}(e_{K})-\omega_{IK}{}^{L}g_{LJ}-\omega_{JK}{}^{L}g_{IL}\,.
\end{split}
\label{Non-metricity tensor in component-form in MAG}
\end{equation}
Considering the covariant exterior derivative of it, the following identity is derived:
\begin{equation}
    d_{\tilde{\nabla}}\tilde{Q}=\tilde{R}\theta^{I}\wedge ge_{I}\,.
\label{}
\end{equation}
This identity is sometimes called zero-th Bianchi's identity.

\bibliography{bibliography.bib}
\bibliographystyle{unsrt}
\end{document}